\documentclass{aa}

\bibpunct{(}{)}{;}{a}{}{,} 

\usepackage[varg]{txfonts}
\usepackage{graphicx}
\usepackage{color}

\begin{document}

\title{A revised estimate of the distance to the clouds in the Chamaeleon complex using the Tycho--Gaia Astrometric Solution}

\author{Jordan Voirin\inst{\ref{inst:ESTEC},\thanks{\email{jordan.voirin@gmail.com}}} \and Carlo F. Manara \inst{\ref{inst:ESTEC}, \thanks{ESA Research Fellow; Current address: European Southern Observatory, Karl-Schwarzschild-Str. 2, D-85748 Garching bei M\"unchen, Germany, \email{cmanara@eso.org}} }  \and Timo Prusti \inst{\ref{inst:ESTEC}}}

\institute{Scientific Support Office, Directorate of Science, European Space Research and Technology Center (ESA/ESTEC), Keplerlaan 1,
2201 AZ Noordwijk, The Netherlands \label{inst:ESTEC}}

\abstract{The determination of the distance to dark star-forming clouds is a key parameter to derive the properties of the cloud itself, and of its stellar content. This parameter is still loosely constrained even in nearby star-forming regions. }
{We want to determine the distances to the clouds in the Chamaeleon-Musca complex and to explore the connection between these clouds and the large scale cloud structures in the galaxy.}
{We use the newly estimated distances obtained from the parallaxes measured by the Gaia satellite and included in the Tycho--Gaia Astrometric Solution catalog. When known members of a region are included in this catalog we use their parallaxes to infer the distance to the cloud. Otherwise, we analyze the dependence of the color excess on the distance of the stars and look for a turn-on of this excess, which is a proxy of the position of the front-edge of the star-forming cloud.}
{We are able to measure the distance to the three Chamaeleon clouds. The distance to Chamaeleon I is $\rm 179^{+11+11}_{-10-10} \textrm{pc}$, where the quoted uncertainties are statistical and systematic uncertainties, respectively, $\sim$20 pc further away than previously assumed. The Chamaeleon II cloud is located at the distance of $\rm 181^{+6+11}_{-5-10} \textrm{pc}$, which agrees with previous estimates. We are able to measure for the first time a distance to the Chamaeleon~III cloud of $\rm 199^{+8+12}_{-7-11} \ \textrm{pc}$. Finally, the distance of the Musca cloud is smaller than 603$^{+91+133}_{-70-92}$ pc. These estimates do not allow us to distinguish between the possibility that the Chamaeleon clouds are part of a sheet of clouds parallel to the galactic plane, or perpendicular to it. }
{We have measured a larger distance to the Chamaeleon~I cloud than assumed in the past, confirmed the distance to the Chamaeleon~II region, and measured for the first time the distance to the Chamaleon~III cloud. These values are consistent with the scenario where the three clouds are part of a single large scale structure. Gaia Data Release 2 will allow us to put more stringent constraints on the distances to these clouds by giving us access to parallax measurements for a larger number of members of these regions.}


\titlerunning{New distances to Chamaeleon complex with TGAS}
\authorrunning{J. Voirin et al.}

   \date{Received May 11, 2017; accepted October 11, 2017}

\maketitle

\section{Introduction}\label{sec:intro}

Star forming regions in the solar neighborhood are ideal laboratories to study pre-main-sequence (PMS) evolution and properties of star-forming clouds. One key parameter in these studies is the distance to the objects. Many physical properties, such as cloud size and mass, stellar luminosity and space motion depend on distance. Any uncertainty in the distance estimate directly translates into uncertainty in the deduced age, mass and other property of Young Stellar Objects (YSOs).

To date, the major limitation in the estimate of the distance to star-forming clouds is the lack of directly measured parallaxes for large number of stars, which are mainly available from the Hipparcos parallaxes \citep{1997A&A...323L..49P} for the brightest nearby stars. Therefore, photometric determined distances have been used in the past \citep[e.g.,][]{Whittet_1997} to derive distances with relatively high uncertainties. 
The Gaia mission \citep{Gaia_2016b} is overcoming this limitation. In particular, the Tycho-Gaia Astrometric Solution (TGAS) from the first Gaia data release  \citep{2016arXiv160904172G} already provides parallaxes to stars down to $G \sim 12$ mag. TGAS data can be used to determine distances both of members of a given star-forming cloud or to field stars in its vicinity. 
Here we use this information to determine the distance to the nearby Chamaleon-Musca star-forming complex.

\begin{figure}
\resizebox{\hsize}{!}{\includegraphics{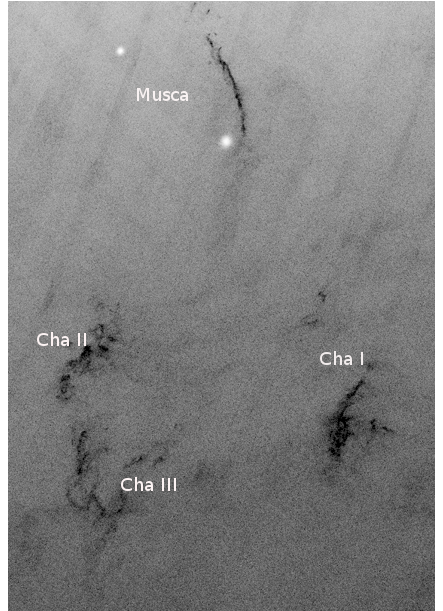}}
\caption{Gaia star density map of the Chamaeleon--Musca region. The figure is a $11^{\circ} \times13^{\circ}$ field centred on the coordinates (l = $300.5^{\circ}$, b = $-13.3^{\circ}$)}
\label{fig:Gaia_Cham_region}
\end{figure}

The proximity of the Chamaeleon--Musca star forming complex has triggered many studies of the system and its PMS population. The complex contains four dark clouds: Chamaeleon I, II, III, and Musca \citep{Luhman_2008}. These objects are visible in the Gaia stellar density map shown in Fig. \ref{fig:Gaia_Cham_region}. The extinction of the central parts of the clouds is so high that very few stars can be seen through to the limiting magnitude $G<20.7$ of Gaia. The complex is located southwards of the Galactic plane around l = 300$^{\circ}$, b = $-15^{\circ}$.

The distance to the Chamaeleon I dark cloud has been extensively discussed in the literature. The previous results vary between 115 and 215 pc \citep{1992lmsf.book...93S}. In \citet{Whittet_1987}, the analysis of the reddening of field stars along the line of sight led to a distance estimate of 140 $\pm$ 12 pc. In \citet{Franco_1991} the distance to the cloud is also found to be 140 pc. The latest result by \citet{Whittet_1997} combined the method of reddening turn-on with other distance indicators and deduced 160 $\pm$ 15 pc. This is the commonly assumed distance to the stars in this region \citep[e.g.][]{Luhman_2008}.

The distance to Chamaeleon II is more difficult to pin down and consequently less certain. Initially, this was constrained to be within 115 pc and 400 pc \citep{Grahm88}. In \citet{Franco_1991} the analysis of the extinction distribution along the line of sight gave a distance of 158 $\pm$ 40 pc. \citet{1992AJ....104..680H} deduced a distance of 200 $\pm$ 20 pc. In \cite{Whittet_1997}, the distance to the Chamaeleon II cloud was found to be 178 $\pm$ 18 pc. 

For the Chamaeleon III dark cloud the distance is poorly constrained in the literature. \citet{Whittet_1997} gives an estimate using the assumption that the Chamaeleon star forming region is on a sheet of clouds parallel to the galactic plane (as later discussed in Sect. \ref{sec:geo_sheet}). Using this assumption their estimate was 140--160 pc. 

The Musca cloud is located northwards of the Chamaeleon region (see Fig. \ref{fig:Gaia_Cham_region}). It is a thin elongated dark cloud possibly about to start forming stars \citep{2016A&A...590A.110C}. Its distance is not well constrained in the literature partly due to the fact that no PMS stars have been identified at optical wavelengths in the cloud. \cite{Franco_1991} argues that it is part of the Chamaeleon system and thus should be around 140 pc.


The paper is structured in five sections. After the introduction, Sect. \ref{sec:method} presents the methods used to determine the distances to the clouds. In Sect. \ref{sec:Results} results are given individually for each cloud. Sect. \ref{sec:Discussion} addresses the question of the physical structure of the Chamaeleon--Musca complex as well as the impact on PMS properties of the newly deduced distances. Finally, in Sect. \ref{sec:Conclusion} we summarize our conclusions.

\section{Methods to determine cloud distances}\label{sec:method}
Two different methods are used to determine the distances to the clouds in the Chamaeleon-Musca complex, and we describe both methods in this section.

\subsection{Members distance}

The distance to a single star may be unreliable as an indication of the distance to a cluster of young stars. Taking advantage of the fact that stars often form together and analyzing distances to all members of a cluster it is possible to have a more reliable estimate. Photometric distances to PMS stars are uncertain as standard colors, extinction and luminosity are all poorly known. By far a more reliable way is to have a direct trigonometric measurement.

We use PMS stars for which the parallax has been directly measured with Gaia. At this stage this is limited to the TGAS catalog \citep{2016arXiv160904172G}. The trigonometric parallaxes have been converted to distances by assuming an anisotropic prior derived from the observed stars in a Milky Way model as discussed in \citet{AABJ}, using the values obtained without including the systematic uncertainties on the parallaxes. In the following we adopt the mode value of the inferred distances plus or minus the difference to the 5 and 95 percentile of the posteriors, following \citet{BJ15}. This implies that the uncertainties we quote on the TGAS distances correspond to a 2$\sigma$ uncertainty.

The distance to a region determined using the distances to individual confirmed members is then taken as the inverse of the mean value of the parallax to the members, weighted for their squared 1$\sigma$ uncertainty, following e.g., \citet{1999AJ....117..354D}.

\subsection{Reddening turn-on distance}\label{sect::met_red}

The second method is based on the principle that the presence of a cloud results in an sudden increase of the extinction at the distance of the front edge of the cloud. By using the measurements of field stars in the line of sight toward a cloud, it is possible to deduce a distance where the increase of the extinction happens. This method is particularly useful in the Chamaeleon complex, given its vicinity and apparent lack of material in front of the clouds \citep{Whittet_1997,1997A&A...326.1215C}. Indeed, this method has limitations if applied to non-homogeneous clouds, where a background star could be located on the line-of-sight of the cloud but observed at negligible extinction, or in cases where significant extinction is present in front of the cloud.

We use only stars for which the parallax has been directly measured with Gaia \citep{Gaia_2016b}. We select the stars spatially located in the line of sight to the clouds. The cloud extent is determined using the Infrared Astronomical Satellites (IRAS) 100$\mu$m flux map\footnote{The maps are taken from the NASA/IPAC Infrared Science Archive (IRSA) website \url{http://irsa.ipac.caltech.edu/applications/DUST/}} \citep{1998ApJ...500..525S}.

The $B-V$ color excess ($\textrm{E}_{(B-V)}$) caused by extinction, is given by the following formula:
\begin{equation}
E_{(B-V)} = (B-V)_{obs} - (B-V)_{int},
\label{eqt:extinction}
\end{equation}
where $(B-V)_{obs}$ is the observed color of a star in the Johnson--Cousins system and $(B-V)_{int}$ is the intrinsic color in the Johnson--Cousins system.

\begin{figure}
\resizebox{\hsize}{!}{\includegraphics{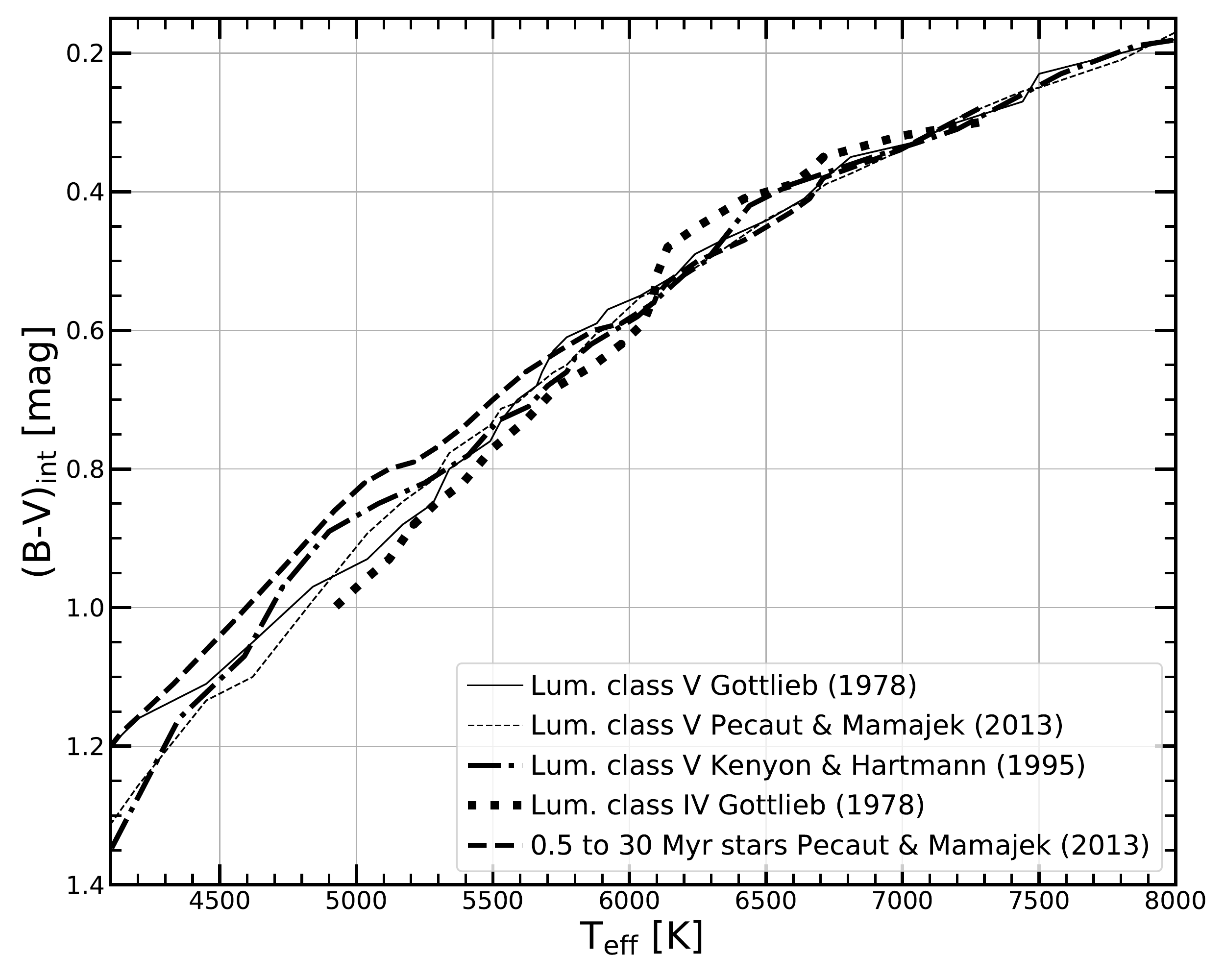}}
\caption{Comparison of $\rm (B-V)_{int}$ to $\rm T_{eff}$ relations for dwarfs and sub-giants.}
\label{fig:comp_int_c}
\end{figure}

The observed colors are computed from the Tycho 2 photometry in the $B_T$ and $V_T$ bands \citep{2000A&A...355L..27H} by transforming to the Johnson--Cousins system as discussed in \citet{1997A&A...323L..49P}. The intrinsic colors depend on the spectral type (SpT) and the luminosity class of the star. Here, we use the relations between the spectral type and the intrinsic color by \citet{Gottlieb_1978} and \citet{2013ApJS..208....9P}. We use the compilation by \citet{2013ApJS..208....9P} for dwarfs and sub-giants with spectral type between K1 and M5, and the one by \citet{Gottlieb_1978} for earlier type sub-giants, for giants, and for stars with luminosity class I and II. The consistency of the two catalogs is shown in Fig.\ref{fig:comp_int_c}. The difference between the two is in general small and always $\le$ 0.1 magnitude for dwarfs and $\le$ 0.15 magnitude for sub-giants. 

The spectral types for the stars in our sample are taken from the reference reported in Simbad \citep{2000A&AS..143....9W}, and these are reported in Tables \ref{tab:ChaI_stars}, \ref{tab:ChaII_stars}, \ref{tab:ChaIII_stars}, and \ref{tab:Musca_stars}. When missing, we determine the luminosity class of the stars using a color magnitude diagram (CMD), as discussed in Appendix \ref{ap:CMDs}. There are stars for which Simbad reports only the effective temperature  ($\rm T_{eff}$)\footnote{The effective temperatures for the stars with no spectral type are either from \cite{2013AJ....146..134K} or \cite{2006ApJ...638.1004A}. In \cite{2013AJ....146..134K} the effective temperatures are deduced from spectroscopy using a combination of two different algorithms, a projection method called MATISSE \citep{2006MNRAS.370..141R} and a decision-tree method called DEGAS \citep{bijaoui:hal-00697630}. The spectra of the stars are fitted to synthesized MARCS model atmospheres \citep{Gustafsson_2008}. In \citet{2006ApJ...638.1004A} the effective temperatures are photometrically computed using a $\chi^2$-optimized fitting polynomials method. \citet{1998A&A...331..619P} and \citet{1999A&AS..140..261A} use also photometric determination by utilizing a minimization method.}. We determine their spectral type using the $\rm T_{eff}$-SpT relation corresponding to the luminosity class determined from the CMD. The $\rm T_{eff}$-SpT relations are taken from \citet{2013ApJS..208....9P} for dwarfs and sub-giants, while the one for giant stars is a combination of \citet{1998A&A...331..619P} and \citet{1999A&AS..140..261A}.

The uncertainty on the color excess is computed using the following formula,
\begin{equation}
\sigma_{E_{(B-V)}} = \sqrt{\sigma^2_{(B-V)_{obs}}+\sigma^2_{(B-V)_{int}}},
\end{equation}
where $\sigma_{E_{(B-V)}}$ is the uncertainty on color excess, $\sigma_{(B-V)_{obs}}$ is the uncertainty on the observed color and $\sigma_{(B-V)_{int}}$ is the uncertainty on the intrinsic color. The uncertainty on the observed color depends on the uncertainty of the Tycho-2 magnitude measurements,
\begin{equation}
\sigma_{(B-V)_{obs}} = 0.85 * \sqrt{\sigma_{B_t}^2+\sigma_{V_t}^2},
\label{eq:BV_Obs}
\end{equation}
where $\sigma_{B_t}$ is the uncertainty on $B_t$ magnitude, $\sigma_{V_t}$ is the uncertainty on $V_t$ magnitude and the factor comes from the transformation between the instrumental magnitude system and the Johnson--Cousins system. The uncertainty on the intrinsic color is directly linked to the spectral type uncertainty. The latter is assumed to be of one sub-class consistent with other work \citep{2011A&A...534A..19A,2006A&A...460..695T}. In the case of stars with only the effective temperature, its uncertainty comprises the uncertainty on the spectral type. For estimating the uncertainty on the intrinsic color the uncertainty of the spectral type is used.

The distance to the cloud is derived from the analysis of the extinction distribution along the line of sight toward field stars. The absorption due to the cloud material causes an increase of the color excess indicating the onset of extinction due to the cloud. The front edge of the cloud lies in the interval between the non-reddened field stars in front of the cloud and the reddened stars behind it. The distance to the cloud is defined as the inverse of the weighted mean parallax of the last non-reddened on-cloud star and the first reddened star. Practically, it means that the star in front of the edge of the cloud must have $\textrm{E}_{(B-V)} + 3\sigma_{\textrm{E}_{(B-V)}} \le 0.2$ mag and the star after the edge must have a $\textrm{E}_{(B-V)} - 3\sigma_{\textrm{E}_{(B-V)}} \ge 0$ mag to be included in the analysis of the cloud distance. The choice of 0.2 mag is driven by the fact that our measurements on not reddened stars lead to negative value of the colors as large as $\sim$-0.2 mag, which is then considered as our systematic uncertainty.


\section{Results}\label{sec:Results}

We present the results for each cloud independently. For the Chamaeleon I cloud both analysis methods could be used, since eight of its confirmed members are included in the TGAS catalog. On the contrary, only the reddening turn-on method can be used to constrain the distance of the other clouds.

\subsection{Chamaeleon I}
\label{sec:ChaI}

\subsubsection{Members analysis}
\label{ap:discr_ChaI}

\begin{figure}
\resizebox{\hsize}{!}{\includegraphics{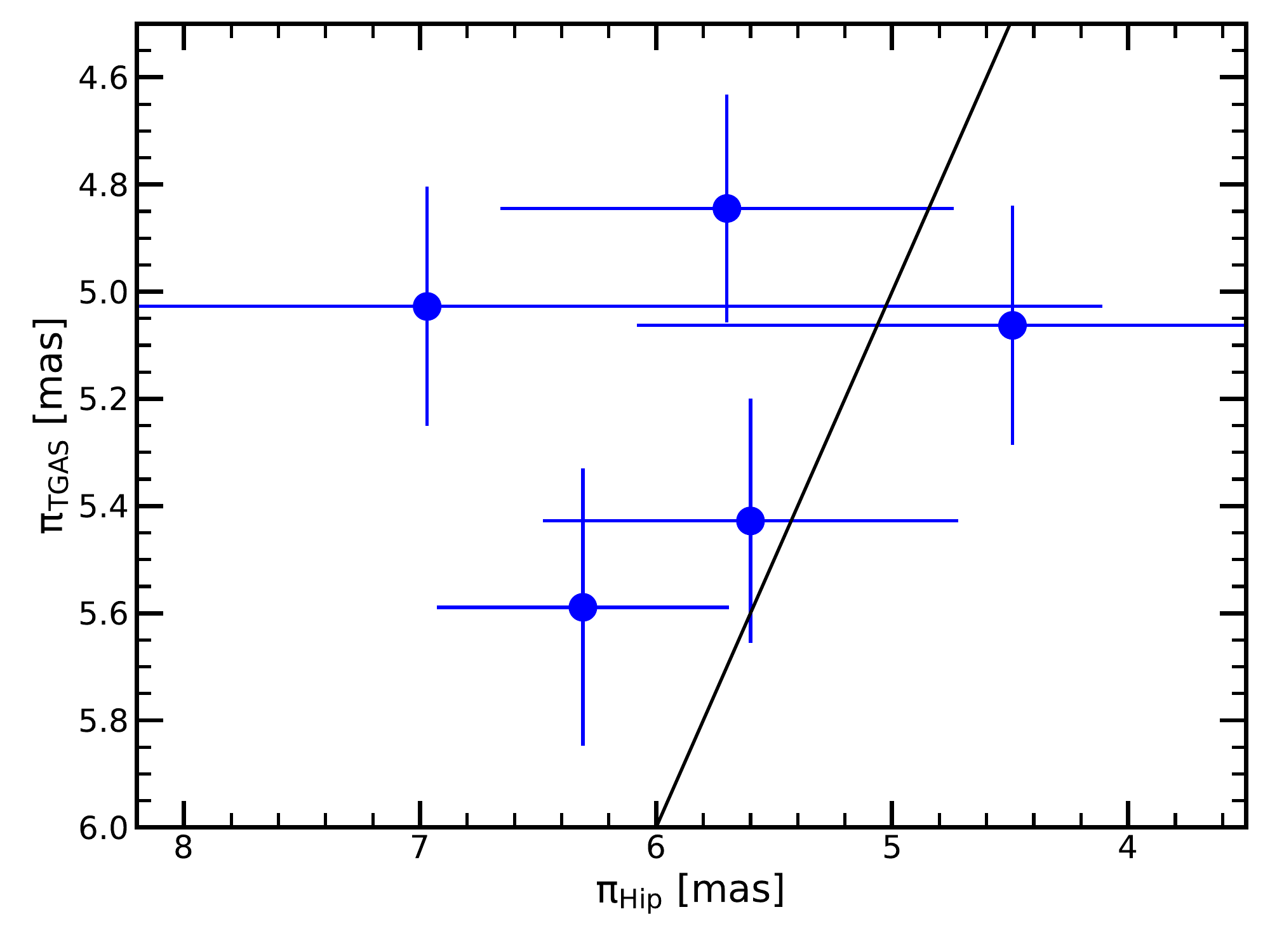}}
\resizebox{\hsize}{!}{\includegraphics{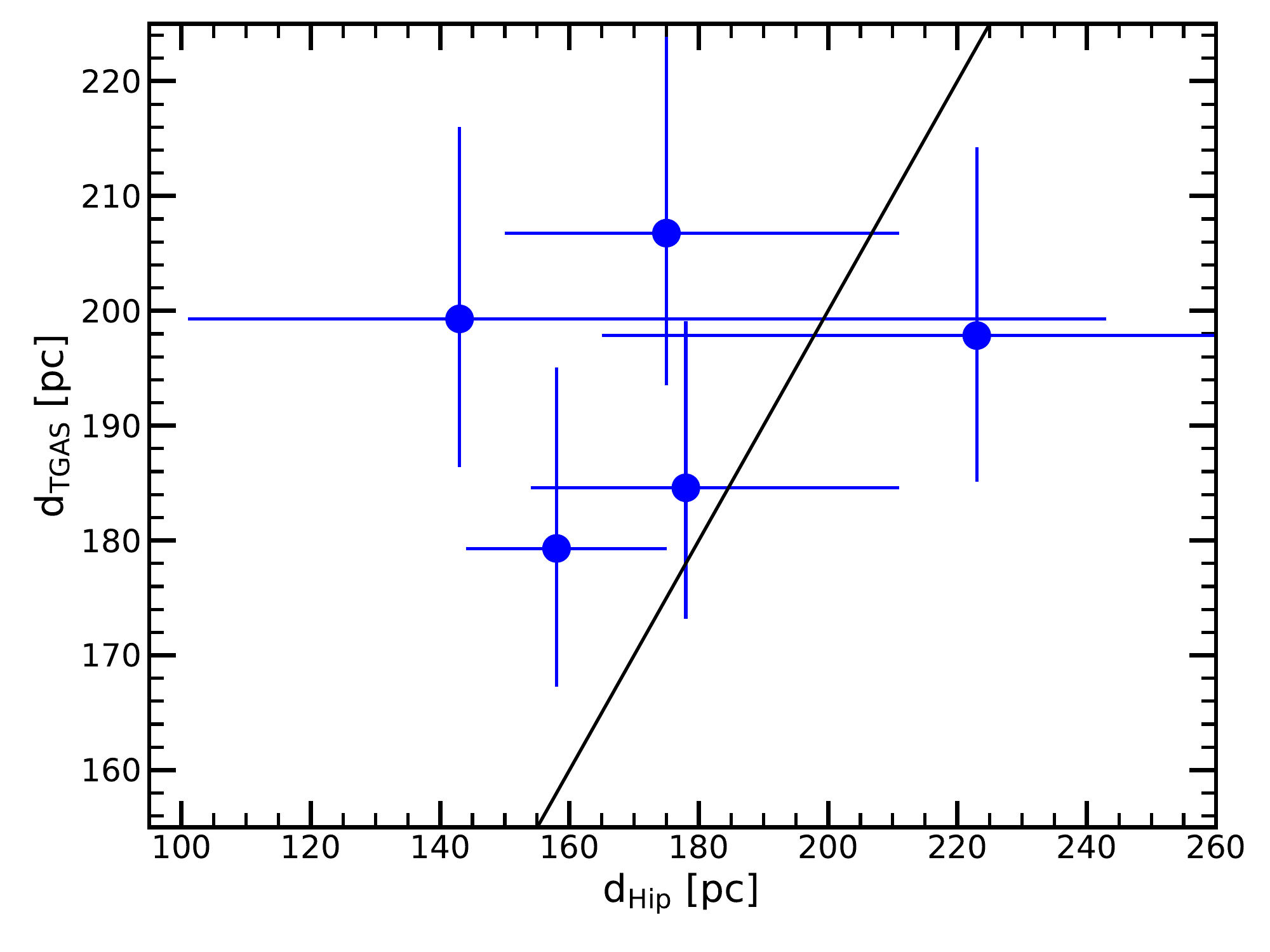}}
\caption{\textit{Top:} TGAS parallaxes ($\mathrm{\pi_{TGAS}}$) vs. Hipparcos parallaxes ($\mathrm{\pi_{Hip}}$) to the five members of Chamaeleon I included in both Hipparcos catalog \citep{2007A&A...474..653V} and TGAS catalog \citep{2016arXiv160904172G}. The black line represents the one to one relation. \textit{Bottom:} Same for distances, taken as inverse of the parallax for Hipparcos data and from \citet{AABJ} for TGAS (see Table~\ref{tab:d_Hip_d_TGAS}). Uncertainties on both axes are as discussed in Table~\ref{tab:d_Hip_d_TGAS}, and they correspond to 1$\sigma$ uncertainty for the Hipparcos values, and 2$\sigma$ for TGAS.
 }
\label{fig:d_Hip_TGAS_comp}
\end{figure}

The Chamaeleon I star-forming region comprises more than 200 members \citep{2004ApJ...602..816L,Luhman_2007,Luhman&Muench_2008}. 
We have searched in the TGAS catalog \citep{2016arXiv160904172G} for members of this region and found that eight of the brightest objects are included. The limited number of matches is due to the magnitude limit ($G\sim$12 mag) of the TGAS catalog. Five of these targets were already included in the Hipparcos catalog \citep{2007A&A...474..653V}. Their parallaxes and distances are reported in Table \ref{tab:d_Hip_d_TGAS}, obtained inverting the parallax for the Hipparcos data and from \citet{AABJ} for TGAS data, and including the errors on these values. While for parallaxes the 1$\sigma$ uncertainty is reported, for distances we compute the 1$\sigma$ asymmetric error on the Hipparcos distances from the inversion of the parallax plus or minus its error, and we report the difference to the 5 and 95 percentile of the median of the TGAS distance reported by \citet{AABJ}. We note that the latter is a 2$\sigma$ uncertainty.  The comparison between the parallaxes and the corresponding inferred distances reported in the two catalogs is shown in Fig. \ref{fig:d_Hip_TGAS_comp}. 
The new estimates of the parallaxes and their inferred distances to these five objects with TGAS are larger than the Hipparcos ones in four cases, although the two sets of values are compatible within their uncertainty.


\begin{table*}
\center
\caption{Hipparcos and TGAS parallaxes and inferred distances of the five members of Chamaeleon I included in both catalogs.}
\label{tab:d_Hip_d_TGAS}
\begin{tabular}{lll|cc|cc}
\hline\hline
Tycho 2 Id & HIP & Gaia DR1 Id & $\pi_{\rm Hip}$ [mas] & $\mathrm{d_{Hip}}$ [pc] & $\pi_{\rm TGAS}$ [mas] &  $\mathrm{d_{TGAS}}$ [pc] \\\hline
TYC 9414-33-1 & 52712 & 5200988654226855680 & 5.70$\pm$0.96 & 175 $^{+36}_{-25}$ & 4.84$\pm$0.21 & 207 $^{+17}_{-13}$ \\ 
... & 54365 & 5201129116838179328 & 4.49$\pm$1.59 & 223$^{+122}_{-58}$ & 5.06$\pm$0.22 & 198$^{+16}_{-13}$ \\
TYC 9414-795-1 & 54413 &5201128120405767040 & 6.31$\pm$0.62 & 158 $^{+17}_{-14}$ & 5.59$\pm$0.26 & 179 $^{+16}_{-12}$ \\ 
TYC 9410-2805-1 & 54557 & 5201344071360118016 & 5.60$\pm$0.88 & 178 $^{+33}_{-24}$ & 5.43$\pm$0.23 & 185 $^{+14}_{-11}$ \\ 
TYC 9410-60-1 & 54744 & 5201295314892489728 & 6.97$\pm$2.86& 143 $^{+100}_{-42}$ & 5.03$\pm$0.22 & 199$^{+17}_{-13}$ \\ 
\hline
\end{tabular}
\tablefoot{Parallaxes from the Hipparcos mission are taken from \citet{2007A&A...474..653V}, and converted to distances as the inverse of the parallax, with the errors calculated using the inverted parallax plus or minus the 1$\sigma$ uncertainty. Parallaxes of Gaia DR1 (TGAS) are taken from \citet{2016arXiv160904172G}. Distances are the mode of the posteriors obtained converting the trigonometric parallaxes assuming an anisotropic prior \citep[Table~3, without systematic uncertainties of][]{AABJ}, and the uncertainties are the difference between the 5th and 95th percentile to the mode of the posterior.}
\end{table*}





\begin{figure}
\resizebox{\hsize}{!}{\includegraphics{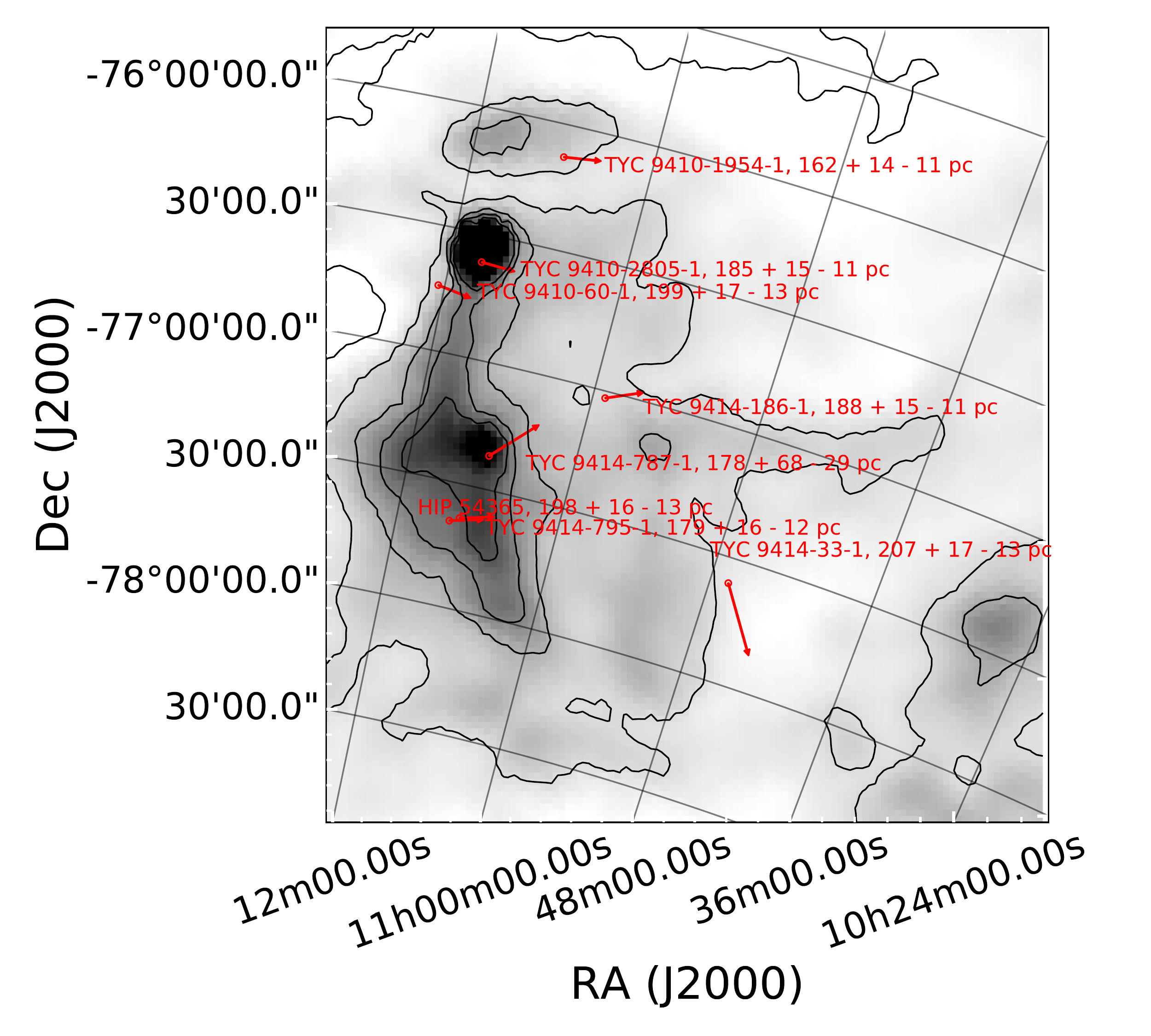}}
\caption{IRAS maps of the Chamaeleon I region at 100 $\mu$m \citep{1998ApJ...500..525S}. The six contours range from 8MJy/sr to 30 MJy/sr. The positions of the members of this region included in the TGAS catalog are represented with red circles. The red arrows represent their displacement in $10^5$ years. The Tycho-2 or Hipparcos name and TGAS distance of each star are reported as labels for each object.}
\label{fig:scatter_ChaI_Vt}
\end{figure}

The positions and proper motions vectors for the eight previously reported members are shown in Fig.\ \ref{fig:scatter_ChaI_Vt}. These objects are located in different parts of the cloud, spanning from the dust cloudlet on the north end to the dust-depleted region to the south, and their individual distances range from $\sim$160 pc to $\sim$200 pc, with typical uncertainties of 8-9 pc. There is a hint of a north-south distance gradient, with the target TYC 9410-1954-1, located in the northern cloudlet, having a distance of 162$^{+14}_{-11}$ pc, stars in the dense region at the center having distances between $\sim$180-190 pc, and the star on the southern part, TYC 9414-33-1 having a distance of 207$^{+17}_{-13}$ pc. However, this could also indicate that the north cloudlet is separated from the main Chamaeleon I cloud, and lays $\sim$ 20 pc closer to us, or even that the Chamaeleon~I cloud is actually composed by three different regions. We are not yet in a position to discriminate between these hypotheses due to the still large uncertainties on the distances and to the limited number of targets. Therefore, we will consider the objects in this cloud to be located at a single distance.

To determine the distance of the Chamaeleon~I cloud based on the distances to its individual members we first note that two of these members have different proper motions and distances with respect to the other members. These are TYC 9414-33-1, which is located outside the cloud on the southern side at a distance of almost 210 pc, and TYC 9410-1954-1, which is at $\sim$160 pc and in the northern cloudlet. The former is considered a member of Chamaelon~I since it has a proper motion direction similar to other members of this region which are too faint to be in the TGAS catalog \citep{Lopez_2013}. We determine the distance to the Chamaeleon~I cloud by performing a weighted mean over the parallaxes to the other six members, and we invert this mean parallax to obtain:
\begin{equation}
\rm \bar{d}_{\rm Chamaeleon \ I, \ members} = 188 ^{+6+11}_{-6-10} \ \rm [pc], 
\label{dist_memb}
\end{equation}
where the uncertainties are 1$\rm \sigma$ statistical and systematic\footnote{All TGAS parallaxes have a systematic uncertainty of 0.3 mas \citep{2016arXiv160904172G}. This is already included in the individual uncertainties on parallaxes in the TGAS catalog, but it affects any mean value, as well, and is therefore always reported in this work.} uncertainties, respectively. This value would be $188 ^{+3+11}_{-3-10}$ pc if TYC 9410-1954-1 is also included, and $189 ^{+3+11}_{-3-10}$ pc if all the eight members included in the TGAS catalog are considered. All these values are consistent within their statistical uncertainties, and we thus use the value reported in Eq.~\ref{dist_memb} as a reference in the following.



\begin{table*}
\caption{\label{tab:ChaI_mem} Members of Chamaeleon I included in the TGAS catalog}
\begin{center}
\begin{tabular}{l ll lll l l}
\hline\hline
Name & Ra (ICRS) & Dec (ICRS) & V$_t$ & B$_t$ & SpT & Ref.&\\\hline
TYC 9414-33-1&161.657371&-77.601034&9.42&10.14&F0V&1&\\ 
TYC 9414-186-1&164.778625&-77.027854&11.79&13.28&K4Ve&4&\\ 
TYC 9410-1954-1&166.490488&-76.130243&7.70&7.87&B6IV/V&1&\\ 
TYC 9414-787-1&166.563573&-77.365761&11.20&13.09&G5Ve&4&...\\ 
HIP 54365&166.835926&-77.635358&11.02&11.84&G2Ve&2&\\ 
TYC 9414-795-1&167.013356&-77.654853&8.49&8.85&A0Vep&4&\\ 
TYC 9410-2805-1&167.458029&-76.613256&9.08&9.48&B9V&1&\\ 
TYC 9410-60-1&168.115046&-76.739532&11.21&12.24&G9Ve&2&\\ 
\hline
\end{tabular}

\begin{tabular}{llllllllllll}
\hline\hline
& $\pi$ [mas] & $\sigma_\pi$ [mas] & $\mu_{\alpha}$ [mas/yr]& $\sigma_{\mu_{\alpha}}$ [mas/yr]&$\mu_{\delta}$ [mas/yr]& $\sigma_{\mu_{\delta}}$ [mas/yr]&E$_{(B-V)}$ & $\sigma_{E_{(B-V)}}$ & d [pc] & $+2\sigma_{d}$ & $-2\sigma_{d}$\\\hline
&4.84&0.21&-28.65&0.15&-9.02&0.13&0.927&0.291&207&+17&-13\\ 
&5.32&0.22&-22.76&0.26&2.21&0.24&0.164&0.393&188&+15&-11\\ 
&6.21&0.28&-22.35&0.04&0.69&0.03&0.411&0.139&162&+14&-11\\ 
...&5.72&0.78&-27.08&3.83&6.04&1.75&0.318&0.144&178&+68&-29\\ 
&5.06&0.22&-22.51&0.20&1.30&0.17&0.093&0.199&198&+16&-13\\ 
&5.59&0.26&-22.46&0.05&1.25&0.05&0.309&0.159&179&+16&-12\\ 
&5.43&0.23&-21.43&0.10&-0.28&0.09&0.050&0.181&185&+15&-11\\ 
&5.03&0.22&-21.69&0.33&-0.86&0.29&0.260&0.079&199&+17&-13\\ 
\hline
\\
\end{tabular}
\tablebib{

- Spectral type reference : (1) \citet{1975mcts.book.....H}; (2) \citet{1977A&A....61...21A};(3) \citet{1977PASP...89..347I}; (4) \citet{2006A&A...460..695T}.

- $V_t$ and $B_t$ are taken from \citet{2000A&A...355L..27H}.

- Proper motions are taken from TGAS \citep{2016arXiv160904172G}.

- Distances are taken from \citet{AABJ} computed with TGAS parallaxes assuming a Milky Way prior. The assumed distance is the mode of the posteriors, and the quoted 2$\sigma$ uncertainties the difference between this value and the 5 and 95 percentile values, respectively.}
\end{center}
\end{table*}

\subsubsection{Reddening turn-on analysis}

\begin{figure}
\resizebox{\hsize}{!}{\includegraphics{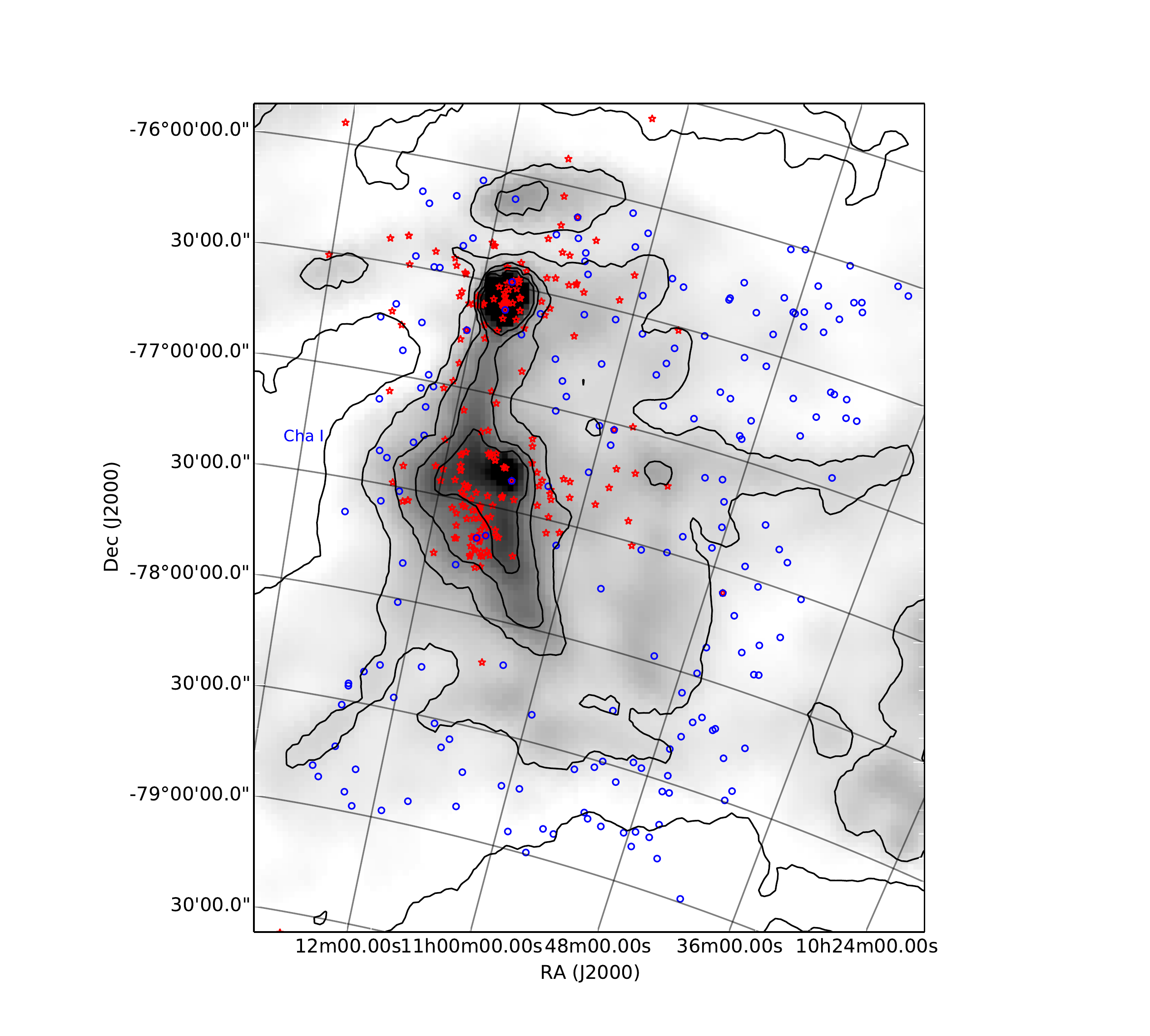}}
\caption{IRAS maps of Chamaeleon I region in 100 $\mu$m flux \citep{1998ApJ...500..525S}. The six contours range from 8MJy/sr to 30 MJy/sr. Red stars represent the known members of Chamaeleon I taken from \citet{Luhman_2007} and \citet{Luhman&Muench_2008}. Blue circles are the TGAS stars selected for our analysis.}
\label{fig:scat_chaI}
\end{figure}

The reddening turn-on method described in Sect.~\ref{sect::met_red} is used here to obtain a distance to the Chamaeleon~I cloud that is independent on the one based on the distances to the members, and also obtained using the distances to more stars, both in or close to the cloud. To determine which stars can be used in this analysis we first compare the position of the members of the Chamaeleon~I region \citep{Luhman_2007,Luhman&Muench_2008} and the dust cloud density map \citep{1998ApJ...500..525S} with the position of all the stars included in the TGAS catalog. We then select only 59 stars from the TGAS catalog that are located on the line-of-sight of the cloud, or in its vicinity. The TGAS stars selected with this criterion are shown in Fig. \ref{fig:scat_chaI}. The number density of stars in the TGAS catalog located on the line-of-sight of the densest parts of the cloud is lower than around the cloud due to the magnitude limit of the TGAS catalog.


\longtab{
\begin{longtable}{llllllllllll}
\caption{\label{tab:ChaI_stars} Catalog of of stars on the line of sight of the Chamaeleon I cloud used in this research}\\
\hline\hline
Name & Ra (ICRS) & Dec (ICRS) & V$_t$ & B$_t$ & SpT & T$_{\rm eff}$ [K] & E$_{(B-V)}$ & $\sigma_{E_{(B-V)}}$ & d [pc] & $\sigma_{d}$ & Ref.\\
\endfirsthead
\caption{continued.}\\
\hline\hline
Name & Ra (ICRS) & Dec (ICRS) & V$_t$ & B$_t$ & SpT & T$_{\rm eff}$ [K] & E$_{(B-V)}$ & $\sigma_{E_{(B-V)}}$ & d [pc] & $\sigma_{d}$ & Ref.\\
\endhead
\hline
\multicolumn{12}{l}{\tablebib{

- Spectral type reference : (1) \citet{1975mcts.book.....H}; (2) \citet{1977A&A....61...21A}; (3) \citet{1984ApJ...283..123V}; (4) \citet{1993psc..book.....B};(5) \citet{1995A&AS..110..367N}; (6) \citet{1996yCat.1116....0J}; (7) \citet{2006A&A...460..695T}; (8) \citet{2011A&A...534A..19A}.

- Effective Temperature reference : (9) \citet{2006ApJ...638.1004A}; (10) \citet{2013AJ....146..134K}.

- $V_t$ and $B_t$ are taken from \citet{2000A&A...355L..27H}.

- The distance are taken from \citet{AABJ} based on TGAS.}}
\endfoot
Field Stars&&&&&&&&&&& \\ \hline 
TYC 9410-508-1&160.868216&-76.580778&8.69&9.55&G5V&...&0.050&0.023&63&1&1\\
TYC 9410-46-1&160.981871&-76.708203&11.54&12.16&K2V&5064&-0.363&0.166&162&8&9\\
TYC 9410-2308-1&161.002202&-76.161670&11.83&13.36&K4IV/V&4572&0.194&0.275&165&8&9\\
TYC 9414-456-1&161.255224&-77.246271&8.95&10.29&G8III&...&0.193&0.035&228&13&1\\
TYC 9410-2267-1&161.372321&-76.332288&10.54&11.29&G5&...&-0.038&0.057&139&6&4\\
TYC 9414-1217-1&161.403717&-78.542314&11.71&12.66&G9V&5382&0.032&0.195&190&14&9\\
TYC 9414-1061-1&161.564597&-78.545900&11.79&12.37&K6V&4200&-0.766&0.247&193&13&9\\
TYC 9418-1454-1&161.862506&-78.758663&11.08&11.88&G5&...&0.008&0.086&178&8&6\\
TYC 9414-415-1&162.166119&-77.419907&11.06&11.66&G6V&5613&-0.197&0.101&164&8&9\\
TYC 9414-278-1&162.214613&-77.203890&10.24&10.98&K0&...&-0.184&0.066&130&6&5\\
TYC 9414-28-1&162.255449&-76.907728&10.33&12.27&K4III&...&0.231&0.188&730&315&3\\
TYC 9410-2438-1&162.291829&-76.237246&8.32&8.62&A1IV&...&0.233&0.044&252&18&1\\
TYC 9414-224-1&162.380665&-77.109418&9.23&9.81&F8IV/V&...&-0.031&0.027&162&6&1\\
TYC 9414-926-1&162.646985&-78.568704&11.78&12.67&K1V&5146&-0.092&0.207&150&7&9\\
TYC 9414-219-1&162.653363&-77.968127&9.94&10.55&B8V&...&0.629&0.039&447&58&3\\
TYC 9414-48-1&162.732705&-77.124600&8.06&8.51&B4V&...&0.545&0.018&266&24&1\\
TYC 9414-707-1&162.816366&-77.410654&11.18&12.17&G2V&5811&0.192&0.133&156&7&9\\
TYC 9414-99-1&163.572856&-77.521715&10.02&10.45&F8&...&-0.170&0.041&143&12&4\\
TYC 9410-368-1&163.952335&-76.864033&11.09&11.76&A2V&...&0.490&0.089&477&179&3\\
TYC 9410-438-1&164.047545&-76.598003&10.42&10.69&B5V&...&0.382&0.048&481&62&3\\
TYC 9414-292-1&164.766076&-77.099141&9.68&11.64&K0III&...&0.670&0.119&207&15&3\\
TYC 9410-2587-1&164.915561&-76.404301&11.22&11.83&G2V&...&-0.136&0.104&157&8&8\\
TYC 9414-178-1&165.088549&-77.026554&10.73&11.60&G5&...&0.052&0.077&139&14&6\\
TYC 9410-2763-1&165.308479&-76.540742&10.71&11.35&B9V&...&0.615&0.074&541&98&3\\
TYC 9414-193-1&165.332242&-77.603393&9.24&9.88&F8V&...&0.011&0.027&124&5&1\\
TYC 9410-532-1&165.357392&-76.751956&9.88&10.43&A9V&...&0.212&0.041&186&20&8\\
TYC 9414-642-1&165.797901&-77.351211&11.60&12.90&G2V&...&0.457&0.222&270&30&8\\
TYC 9414-631-1&165.807731&-78.191607&11.17&12.73&K5V&4500&0.195&0.200&83&2&10\\
TYC 9410-2781-1&165.925807&-76.553458&11.71&12.62&K3V&4800&-0.221&0.234&102&3&10\\
TYC 9410-464-1&166.032810&-76.870223&11.56&14.16&K3III&...&0.932&0.435&840&883&3\\
TYC 9410-2362-1&166.042540&-76.371735&10.97&11.82&A7V&...&0.516&0.093&684&220&8\\
TYC 9418-14-1&166.081702&-78.868312&8.95&9.92&K0V&...&0.006&0.031&46&0&1\\
TYC 9410-2504-1&166.158603&-76.317068&11.44&12.18&F8V&...&0.102&0.152&214&13&8\\
TYC 9410-2172-1&166.182226&-76.279181&11.56&12.62&F5V&...&0.458&0.214&859&347&8\\
TYC 9410-496-1&166.272420&-76.780605&11.92&13.25&K1IV/V&5217&0.206&0.300&186&12&9\\
TYC 9410-230-1&166.764576&-76.595474&12.26&13.41&K1V&5155&0.124&0.340&184&35&10\\
TYC 9410-2094-1&166.810267&-76.228772&11.47&12.57&G8V&5435&0.198&0.192&130&6&9\\
TYC 9410-408-1&167.038953&-76.707223&10.06&10.70&F7V&6236&0.034&0.043&165&8&9\\
TYC 9414-695-1&167.326314&-77.794525&8.55&9.05&F5V&...&-0.013&0.029&95&3&1\\
TYC 9414-341-1&167.576825&-78.282897&11.22&11.79&K0V&5287&-0.332&0.113&191&11&9\\
TYC 9410-1834-1&167.716180&-76.110897&8.79&9.51&G3V&...&-0.056&0.021&121&9&8\\
TYC 9410-2449-1&168.359348&-76.323246&11.66&12.38&G2V&...&-0.041&0.170&242&26&8\\
TYC 9414-649-1&168.450598&-77.836635&11.73&13.21&G7III&...&0.319&0.302&212&24&8\\
TYC 9414-692-1&168.499507&-78.312784&9.74&10.45&G2V&5761&-0.048&0.039&132&4&10\\
TYC 9414-462-1&168.538915&-77.247064&11.12&13.29&K3III&...&0.565&0.318&633&463&8\\
TYC 9414-260-1&168.546833&-77.020894&11.02&11.62&F2V&...&0.138&0.086&330&34&8\\
TYC 9418-1195-1&168.580385&-78.801584&10.46&11.44&G5&...&0.160&0.063&133&5&4\\
TYC 9414-392-1&168.622063&-77.118579&11.26&12.12&G2V&...&0.082&0.131&180&13&8\\
TYC 9414-46-1&168.687976&-76.972473&11.46&12.89&G4III&...&0.391&0.217&870&443&8\\
TYC 9414-542-1&168.820157&-77.518391&7.67&8.22&B8III&...&0.576&0.034&280&21&1\\
TYC 9410-1825-1&168.822008&-76.149095&11.74&12.09&B9V&...&0.369&0.142&759&444&8\\
TYC 9410-2634-1&168.877424&-76.483964&11.82&13.44&F8V&...&0.842&0.299&253&39&8\\
TYC 9410-2731-1&168.988700&-76.485709&11.50&14.12&M0III&...&0.679&0.366&903&1060&8\\
TYC 9414-849-1&169.150815&-78.715828&10.65&11.55&G7V&5512&0.049&0.064&119&4&9\\
TYC 9410-2612-1&169.378169&-76.452163&10.29&12.68&M3III&...&0.421&0.178&1232&969&8\\
TYC 9418-716-1&169.412021&-78.865635&11.73&13.16&K5V&4472&0.078&0.248&101&6&10\\
TYC 9414-80-1&169.582445&-77.121264&10.10&11.39&G7III&...&0.158&0.060&440&53&8\\
TYC 9410-488-1&169.588064&-76.682217&11.04&13.70&K6III&...&0.758&0.294&1783&1182&8\\
TYC 9418-610-1&169.590124&-78.819131&10.42&10.90&F8V&6167&-0.125&0.046&176&14&10\\ 
\end{longtable}
}

Both spectral type and luminosity class are available for all the eight Chamaeleon I members included in the TGAS catalog and for 35 other stars selected for the analysis. In addition, we derive the luminosity class for six field stars in the TGAS catalog missing this information from the literature, and for 18 field stars in the TGAS catalog for which only the effective temperature is estimated in the literature. As shown in Appendix \ref{ap:CMDs}, 22 of these stars are dwarfs and two have luminosity class between sub-giants and dwarfs (see Table \ref{tab:ChaI_stars}).

\begin{figure}
\resizebox{\hsize}{!}{\includegraphics{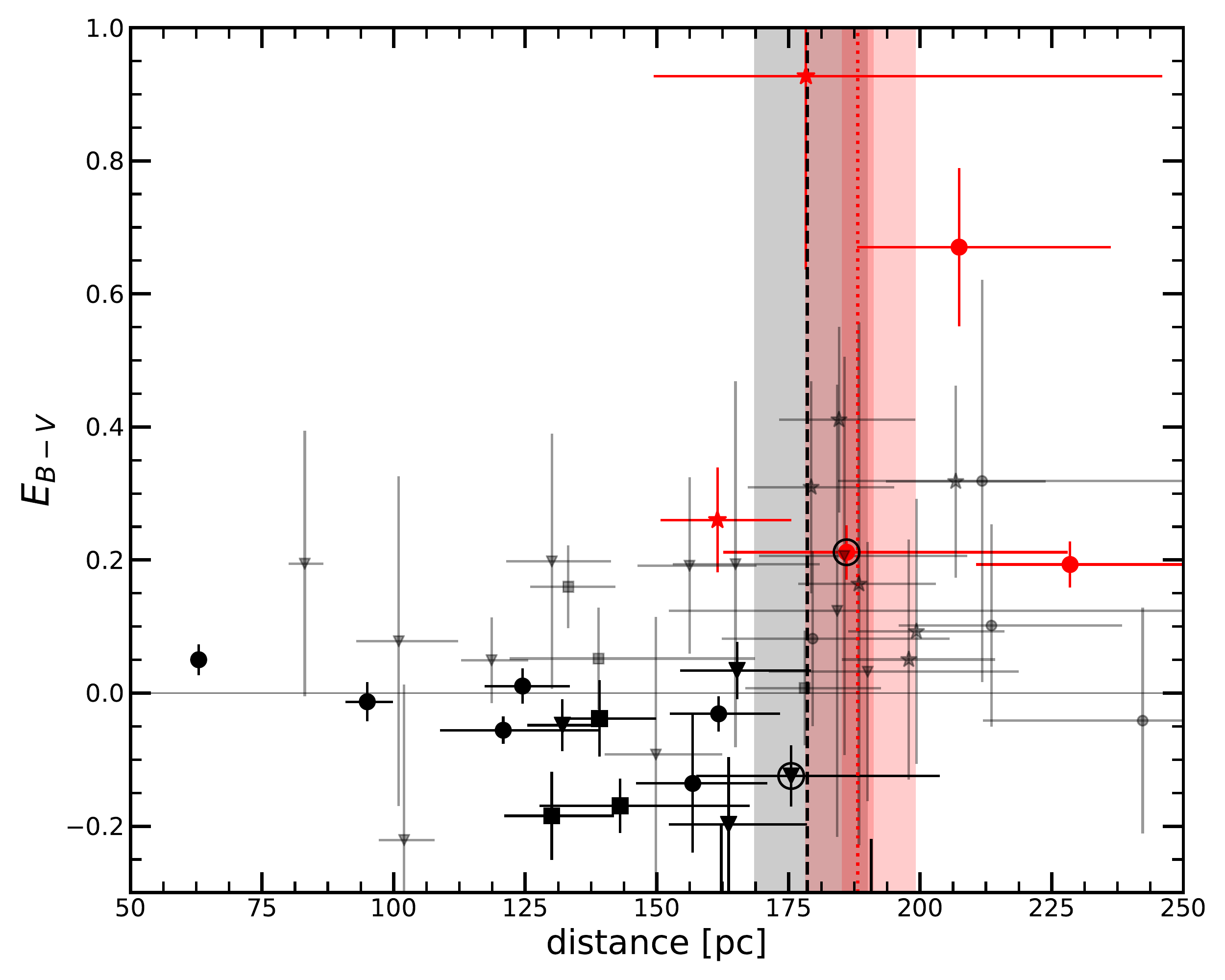}}
\caption{Extinction distribution along the line of sight for stars located on the line-of-sight of the Chamaeleon~I cloud. The dashed line represents the position of the front edge of the cloud determined by the reddening turn-on method, and the gray shaded region its 1$\sigma$ uncertainty. The red dotted line is the average distance of the members, shown here for comparison, and the red shaded region is the 1$\sigma$ uncertainty on this value, either statistical (darker red region) or systematic (lighter red region). The members of the Chamaeleon~I region are represented with stars, the field stars for which both spectral type and luminosity class are reported in the literature are shown with circles, whereas triangles are used for the objects for which we estimate the luminosity class ourselves. Symbols colored in red are used for the stars with color excess larger than 0 at 3$\sigma$, in black with color excess $<$0.2, and in gray when the uncertainties are too large to be considered in either of these two categories. We highlight with black circles the stars used to compute the distance of the reddening turn-on, thus the front edge of the cloud. }
\label{fig:E_BV_d_ChamI}
\end{figure}

We show in Fig. \ref{fig:E_BV_d_ChamI} the measured $E(B-V)$ color excess versus the distance obtained from the TGAS parallax for all the stars. We observe a turn-on of the color-excess, which we consider the distance of the front edge of the cloud, at a parallax of 5.60$\pm$0.34 mas, corresponding to a distance of:
\begin{equation}
\rm d_{Chamaeleon\ I} = 179 ^{+11+11}_{-10-10} \ \textrm{[pc]}, 
\end{equation} 
where the uncertainties are 1$\rm \sigma$ statistical and systematic uncertainties, respectively. The weighted mean parallax is obtained from the measured parallaxes of the two objects used in the computation, and the statistical uncertainty is obtained propagating their individual uncertainties. The distance is derived inverting the parallax, and the statistical and systematic uncertainties by computing the difference between the distance and the inverted parallax plus or minus the statistical or systematic uncertainties. 
This value is compatible within the 1$\sigma$ uncertainties with the distance determined considering only the members of Chameleon~I. This agreement confirms that both methods are viable to derive distances to star forming regions, and we will thus use the reddening turn-on method in the following to measure the distances to the other Chamaeleon clouds, where no members are included in the TGAS catalog. The discrepancy on the nominal values of the distances determined with the two methods is also possibly ascribed to the fact that they probe two different regions of the cloud. The reddening turn-on method measures the distance to the front-edge of the cloud, while the mean distance of the members is sensitive to the radial extent of the cloud itself. Assuming that the cloud is as deep as it is extended in the sky, namely $\sim$ 2$^\circ$, $\sim$ 7 pc at 179 pc, the difference in the nominal values is consistent with this possibility. Since we know that stars form at all positions within the star-forming clouds and that the selection of members of the region included in the TGAS catalog is highly incomplete and biased by the magnitude limit of this catalog, and thus by the intrinsic luminosity of the targets and by the amount of reddening on the line-of-sight, we consider the distance of the front-edge of the cloud as the real distance to the Chamaeleon~I cloud.

\begin{figure}
\resizebox{\hsize}{!}{\includegraphics{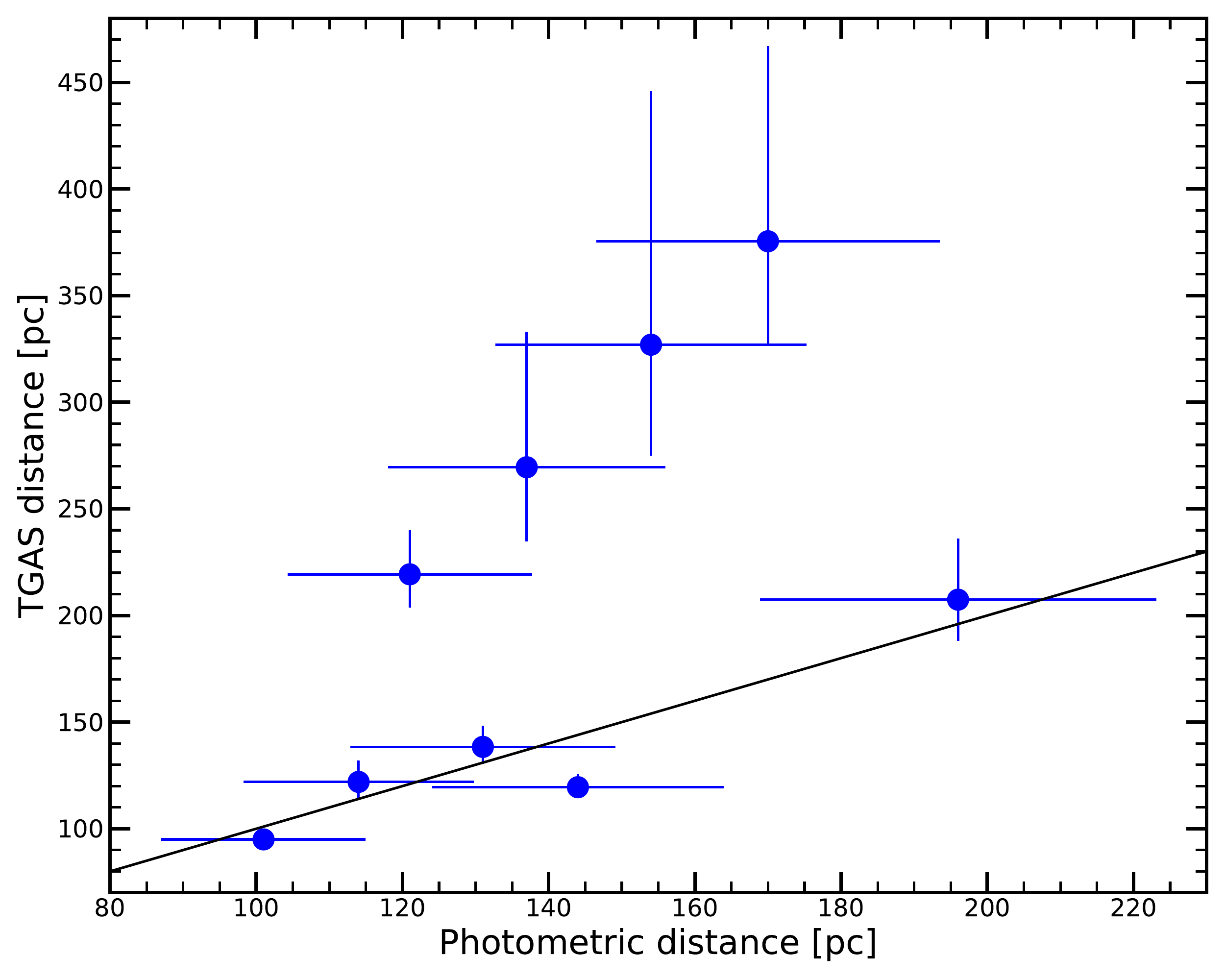}}
\caption{Comparison of photometric distances ($\rm d_{phot}$) vs. TGAS distances ($\rm d_{TGAS}$) for the \cite{Whittet_1997} stars that are relevant for the computation of the onset. The black line represents $\rm d_{TGAS} = d_{\rm phot}$.}
\label{fig:Comp_phot_TGAS_d}
\end{figure}

We note that this newly estimate distance for the Chamaelon~I region is compatible within 1$\sigma$ uncertainties with the previous estimate of 160$\pm$15 pc \citep{Whittet_1997}, but that the nominal value is 19 pc further away than assumed. This discrepancy is due to the fact that the distances derived from the TGAS parallaxes are statistically significantly larger than the photometric distances for the same stars used by \citet{Whittet_1997}, which are reported in Table \ref{tab:comp_dist} and shown in Fig. \ref{fig:Comp_phot_TGAS_d}. Thus, some stars considered to be in front of the cloud by \citet{Whittet_1997} are actually behind the cloud. We discuss that the possible reasons for the difference is an incorrectly estimated luminosity class in the previous works in Appendix \ref{ap:discr_Dist}. 

\vspace{0.7cm}


\subsection{Chamaeleon II}\label{sec:ChaII}

\begin{figure}
\resizebox{\hsize}{!}{\includegraphics{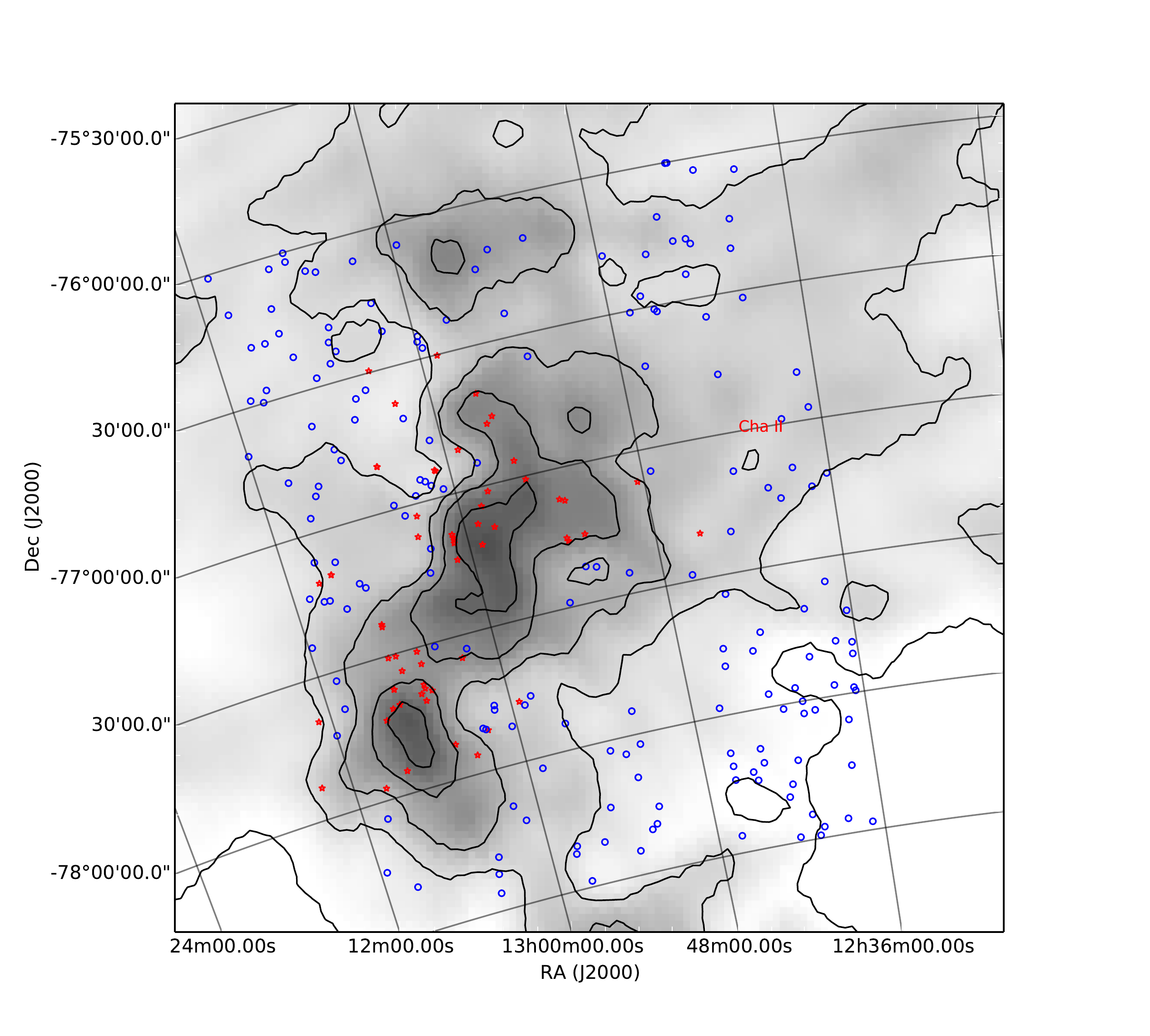}}
\caption{IRAS 100 $\mu$m flux maps of Chamaeleon II region. The six contours range from 8MJy/sr to 30 MJy/sr. Red stars represent the confirmed members of Chamaeleon II reported in \citet{Spezzi_2008}. Blue circles are the TGAS stars in this field of view.}
\label{fig:scat_chaII}
\end{figure}

The distance of the Chamaeleon~II cloud is derived here using solely the reddening turn-on method, since no members of this region are included in the TGAS catalog due to their faintness. Fig. \ref{fig:scat_chaII} shows 
the position of the known members of the Chamaeleon~II region and of the 44 stars included in the TGAS catalog located on the line-of-sight of this region and used for the analysis. 


Both spectral type and luminosity class are available from the literature for 28 of the TGAS stars on the line-of-sight of the Chamaeleon~II cloud. For the remaining stars, the luminosity class is missing in five cases, and only the effective temperature is reported for 11 stars. These luminosity classes are computed as discussed in Appendix \ref{ap:CMDs}, and we find that they are always dwarfs, apart from two cases where they are between sub-giants and dwarfs (see Table \ref{tab:ChaII_stars}).

\longtab{
\begin{longtable}{llllllllllll}
\caption{\label{tab:ChaII_stars} Catalog of of stars on the line of sight of the Chamaeleon II cloud used in this research}\\
\hline\hline
Name & Ra (ICRS) & Dec (ICRS) & V$_t$ & B$_t$ & SpT & T$_{\rm eff}$ [K] & E$_{(B-V)}$ & $\sigma_{E_{(B-V)}}$ & d [pc] & $\sigma_{d}$ & Ref.\\
\endfirsthead
\caption{continued.}\\
\hline\hline
Name & Ra (ICRS) & Dec (ICRS) & V$_t$ & B$_t$ & SpT & T$_{\rm eff}$ [K] & E$_{(B-V)}$ & $\sigma_{E_{(B-V)}}$ & d [pc] & $\sigma_{d}$ & Ref.\\
\endhead
\hline
\multicolumn{12}{l}{\tablebib{

- Spectral type reference : (1) \citet{1975mcts.book.....H}; (2) \citet{1992AJ....104..680H}; (3) \citet{1993psc..book.....B}; (4) \citet{1996yCat.1116....0J}.

- Effective Temperature reference : (4) \citet{2006ApJ...638.1004A}; (5) \citet{2013AJ....146..134K}.

- $V_t$ and $B_t$ are taken from \citet{2000A&A...355L..27H}.

- The distance are taken from \citet{AABJ} based on TGAS.}}
\endfoot
Field Stars&&&&&&&&&&& \\ \hline 
TYC 9416-302-1&189.088504&-77.702072&9.09&9.76&G0V&...&-0.021&0.022&100&3&1\\
TYC 9412-48-1&189.283711&-76.824509&10.40&11.19&G5&...&-0.003&0.057&151&10&3\\
TYC 9412-611-1&189.727451&-76.066041&6.97&9.02&K4III&...&0.320&0.016&216&16&1\\
TYC 9416-279-1&189.841740&-77.846092&9.28&10.54&G6III&...&0.152&0.030&367&40&1\\
TYC 9417-1732-1&190.177962&-77.948498&8.97&10.79&K2III&...&0.380&0.034&547&91&1\\
TYC 9413-1-1&190.517227&-76.785004&10.29&10.94&G0V&5917&-0.044&0.052&164&8&5\\
TYC 9413-2300-1&190.535448&-76.573299&11.84&12.55&K3V&4796&-0.383&0.210&149&22&6\\
TYC 9417-58-1&190.771915&-77.351070&8.68&9.06&B9IV&...&0.370&0.018&301&22&1\\
TYC 9413-1434-1&190.826348&-76.283027&12.21&13.29&K4V&4718&-0.178&0.318&200&19&5\\
TYC 9417-587-1&191.578906&-78.438418&10.82&11.36&F8&...&-0.070&0.075&178&8&3\\
TYC 9413-2539-1&191.614882&-76.707284&11.46&12.39&K6V&4265&-0.466&0.203&200&13&5\\
TYC 9417-1920-1&191.865764&-77.081772&10.79&11.79&F8&...&0.317&0.077&195&25&4\\
TYC 9417-156-1&192.549518&-77.425046&11.19&11.87&K1V&5198&-0.272&0.104&151&6&5\\
TYC 9417-891-1&193.004135&-78.037554&11.45&13.55&G5III&...&0.907&0.292&1113&812&2\\
TYC 9417-1367-1&193.022889&-77.914597&9.85&10.52&F8V&6115&0.044&0.036&121&4&5\\
TYC 9417-560-1&193.061285&-77.378283&11.11&11.74&A5V&...&0.379&0.081&286&22&2\\
TYC 9417-1193-1&193.284001&-78.062156&11.31&12.63&K0III&...&0.123&0.182&480&62&2\\
TYC 9417-87-1&193.421521&-78.413817&10.99&12.03&K4V&4682&-0.220&0.119&65&1&5\\
TYC 9417-877-1&193.539396&-78.037035&7.94&9.43&K0III&...&0.267&0.019&184&8&1\\
TYC 9413-1970-1&193.575886&-76.413166&10.87&11.92&K2IV/V&5071&-0.021&0.093&209&11&5\\
TYC 9413-1821-1&193.602903&-76.175376&9.82&11.06&G8/K0III&...&0.110&0.046&186&11&1\\
TYC 9417-518-1&193.624030&-77.482674&11.32&11.98&A7V&...&0.354&0.106&416&388&2\\
TYC 9417-1320-1&194.754427&-78.039813&12.40&13.00&M1.5V&3647&-0.975&0.358&199&10&5\\
TYC 9417-1726-1&194.780652&-77.802499&11.15&11.96&G2V&5772&0.032&0.097&183&11&5\\
TYC 9413-1675-1&194.909860&-76.081359&10.28&11.27&K0&...&0.023&0.063&56&1&4\\
TYC 9417-1432-1&195.083920&-77.865309&11.50&12.09&F0V&...&0.207&0.127&364&49&2\\
TYC 9417-1915-1&195.172404&-76.974126&10.94&11.78&G1V&...&0.109&0.085&136&5&2\\
TYC 9413-169-1&195.187412&-76.791959&10.88&11.61&G0&...&0.025&0.092&194&11&3\\
TYC 9417-675-1&195.265157&-78.206597&10.51&12.75&K3III&...&0.619&0.178&621&392&2\\
TYC 9417-1924-1&195.493968&-76.920605&11.03&12.53&K1III&...&0.182&0.178&981&537&2\\
TYC 9417-1595-1&195.531262&-77.851905&11.11&12.05&K3V&...&-0.191&0.111&87&2&2\\
TYC 9417-1646-1&195.574668&-77.845509&11.54&12.34&F5III&...&0.220&0.159&415&47&2\\
TYC 9417-1873-1&195.620861&-77.170378&10.37&12.28&K3III&...&0.344&0.138&619&166&2\\
TYC 9417-1119-1&195.986743&-78.368489&11.66&12.61&K0V&5298&-0.009&0.206&149&6&5\\
TYC 9417-698-1&195.989636&-77.514367&12.62&13.64&G1V&...&0.265&0.362&207&18&2\\
TYC 9413-2550-1&196.229021&-76.649512&8.72&9.16&F2IV&...&0.023&0.019&157&9&1\\
TYC 9413-1000-1&196.578609&-76.001823&10.13&10.88&F8/G2V&...&0.048&0.044&178&8&1\\
TYC 9417-232-1&196.891525&-77.221583&9.96&10.38&A0V&...&0.355&0.035&331&46&1\\
TYC 9413-2441-1&196.978742&-76.182897&9.95&10.75&G8/K1V&...&-0.348&0.041&65&1&1\\
TYC 9413-97-1&197.072476&-76.843349&10.42&10.66&A0V&...&0.208&0.045&389&59&1\\
TYC 9417-730-1&197.206239&-77.296039&10.39&12.13&K1III&...&0.377&0.115&610&455&2\\
TYC 9417-1823-1&197.340652&-76.946374&9.98&11.11&G4III&...&0.131&0.047&670&373&2\\
TYC 9417-152-1&197.526954&-77.247556&9.68&9.97&A2III/IV&...&0.202&0.028&281&23&1\\
TYC 9417-1712-1&197.995827&-77.722052&10.45&11.06&F3V&...&0.138&0.054&208&13&2\\ 
\end{longtable}
}

\begin{figure}
\resizebox{\hsize}{!}{\includegraphics{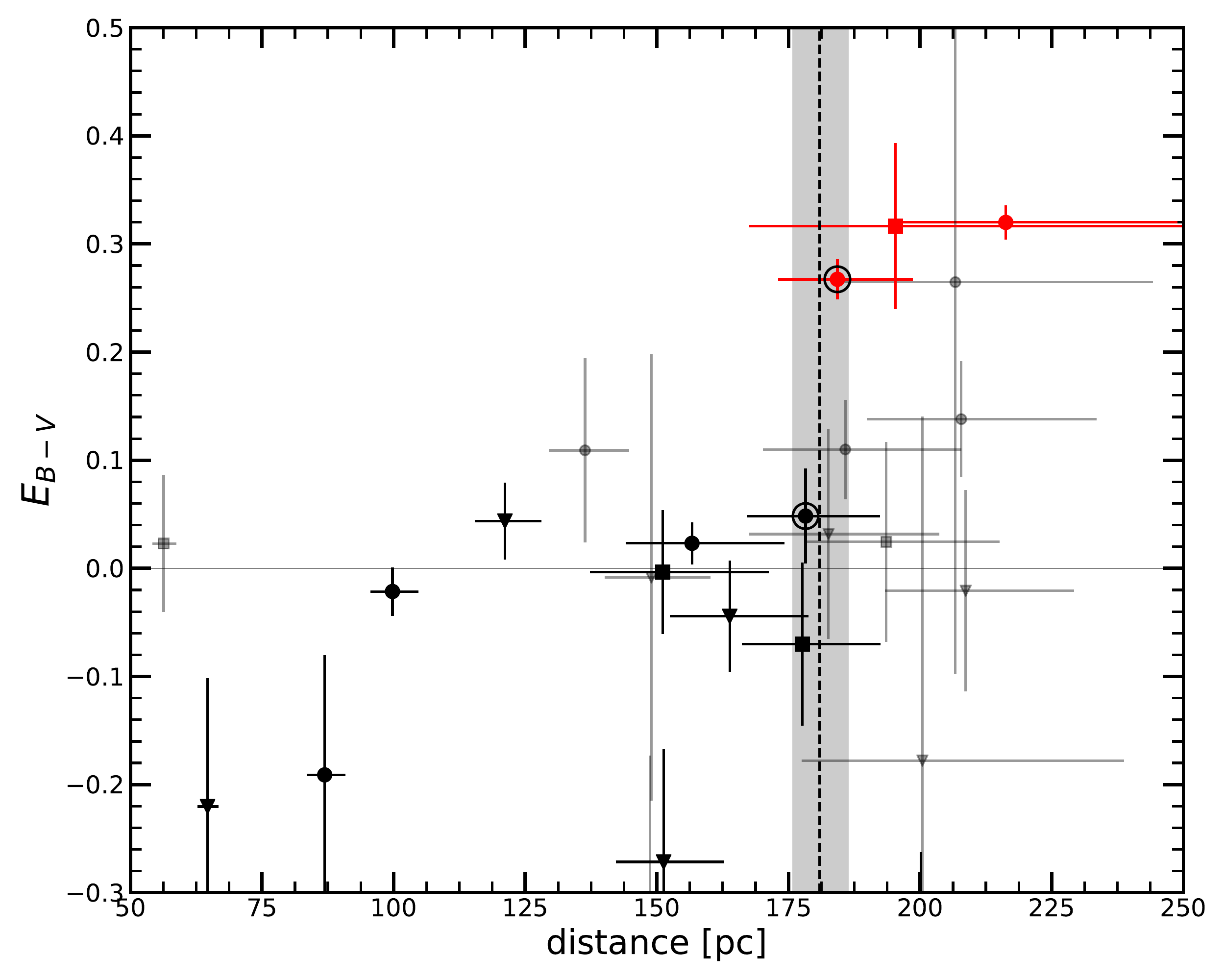}}
\caption{Color excess as a function of distance for stars included in the TGAS catalog and located on the line-of-sight of the Chamaeleon~II cloud. The dashed line represents the position of the cloud front edge, and the gray shaded region its uncertainty. Other symbols are as in Fig.~\ref{fig:E_BV_d_ChamI}.}
\label{fig:E_BV_d_ChamII}
\end{figure}

We thus compute the $E(B-V)$ color excess for these targets, and this is shown as a function of their distance in Fig. \ref{fig:E_BV_d_ChamII}. The distance of the reddening turn-on, and thus of the front-edge of the cloud, is found to be at a parallax of 5.53$\pm$0.16 mas, corresponding to a distance of:
\begin{equation}
\rm d_{Chamaeleon\ II} = 181^{+6+11}_{-5-10} \ \textrm{[pc]}, \nonumber
\end{equation} 
which is well compatible with the previous estimate of 178$\pm$18 pc by \citet{Whittet_1997}, and definitely rules out the hypothesis of much larger distances of 400 pc previously claimed for this cloud. 


\subsection{Chamaeleon III}\label{sec:ChaIII}

\begin{figure}
\resizebox{\hsize}{!}{\includegraphics{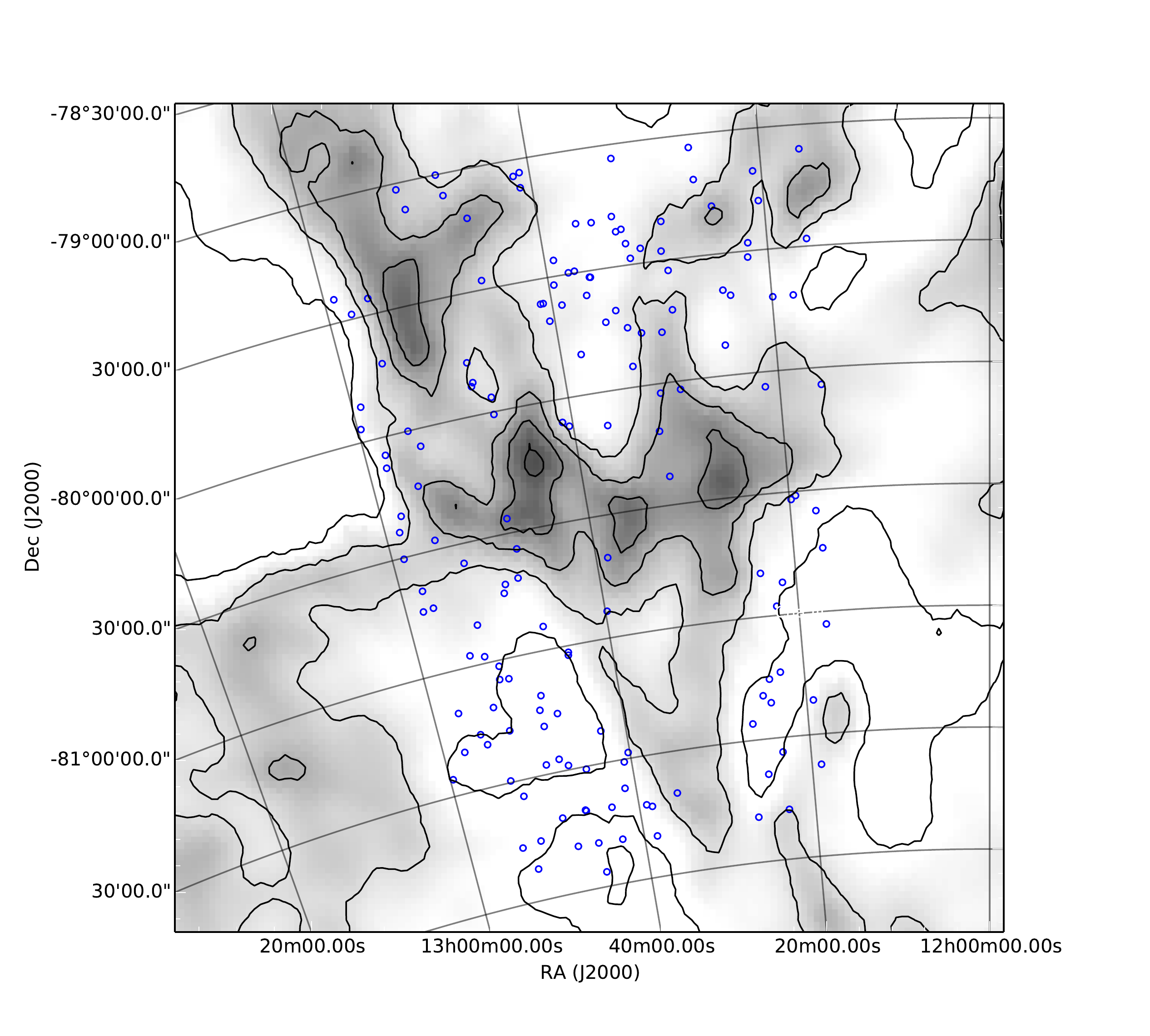}}
\caption{IRAS maps of Chamaeleon III region in 100 $\mu$m flux. The six contours range from 8 MJy/sr to 30 MJy/sr. Blue circles are the TGAS stars in this field of view.}
\label{fig:scat_chaIII}
\end{figure}

The Chamaeleon~III region is less studied than the previous ones, and no information about its young stellar population is available. We thus select for the analysis of its distance the 39 stars included in the TGAS catalog which are located on the line-of-sight of the cloud, as shown in Fig. \ref{fig:scat_chaIII}. 
For only nine of these stars we are able to find in the literature both spectral type and luminosity class information, while we determine the luminosity class for nine additional stars missing this information, and for the remaining 21 stars for which only the effective temperature is reported. As discussed in Appendix \ref{ap:CMDs}, all but five of these stars are main sequence stars, while the others are in between sub-giants and dwarfs in three cases, and sub-giants in the last two cases (see Table \ref{tab:ChaIII_stars}).

\longtab{
\begin{longtable}{llllllllllll}
\caption{\label{tab:ChaIII_stars} Catalog of of stars on the line of sight of the Chamaeleon III cloud used in this research}\\
\hline\hline
Name & Ra (ICRS) & Dec (ICRS) & V$_t$ & B$_t$ & SpT & T$_{\rm eff}$ [K] & E$_{(B-V)}$ & $\sigma_{E_{(B-V)}}$ & d [pc] & $\sigma_{d}$ & Ref.\\
\endfirsthead
\caption{continued.}\\
\hline\hline
Name & Ra (ICRS) & Dec (ICRS) & V$_t$ & B$_t$ & SpT & T$_{\rm eff}$ [K] & E$_{(B-V)}$ & $\sigma_{E_{(B-V)}}$ & d [pc] & $\sigma_{d}$ & Ref.\\
\endhead
\hline
\multicolumn{12}{l}{\tablebib{

- Spectral type reference : (1) \citet{1975mcts.book.....H}; (2) \citet{1993psc..book.....B}; (3) \citet{2006A&A...460..695T}; (4) \citet{1996yCat.1116....0J}.

- Effective Temperature reference : (5) \citet{2006ApJ...638.1004A}; (6) \citet{2013AJ....146..134K}.

- $V_t$ and $B_t$ are taken from \citet{2000A&A...355L..27H}.

- The distance are taken from \citet{AABJ} based on TGAS.}}
\endfoot
 Field Stars&&&&&&&&&&& \\ \hline 
TYC 9420-439-1&184.362157&-80.585270&8.63&9.62&G4III&...&0.010&0.021&268&19&3\\
TYC 9420-1202-1&184.509344&-79.695839&10.79&11.46&F0&...&0.278&0.082&246&21&4\\
TYC 9420-1235-1&185.117516&-79.298279&10.69&11.57&G5&...&0.072&0.075&152&9&4\\
TYC 9420-1121-1&185.436109&-79.466644&7.51&7.55&A1V&...&-0.006&0.016&154&7&1\\
TYC 9420-1483-1&185.468407&-79.525054&9.62&10.00&F0&...&0.028&0.049&170&8&2\\
TYC 9424-756-1&185.568779&-80.962481&10.21&10.88&F8&...&0.040&0.047&122&4&2\\
TYC 9424-37-1&185.763786&-81.796423&11.10&11.75&K4V&4689&-0.548&0.131&217&14&5\\
TYC 9424-743-1&185.783305&-81.559758&12.35&13.75&K3V&4769&0.206&0.312&189&10&5\\
TYC 9424-795-1&185.969374&-81.354016&8.50&8.96&A8IV&...&0.136&0.018&316&27&1\\
TYC 9420-1181-1&186.159243&-79.302519&10.83&11.52&G0&...&-0.012&0.085&177&10&2\\
TYC 9420-1-1&187.826019&-80.039168&11.08&12.51&K1V&5115&0.374&0.178&104&4&5\\
TYC 9420-107-1&187.974705&-80.193543&9.48&10.05&G0&...&-0.106&0.051&106&3&2\\
TYC 9420-339-1&188.076799&-79.783818&7.10&7.18&A0V&...&0.065&0.015&154&9&1\\
TYC 9420-320-1&188.210984&-79.057732&11.87&12.84&G8V&5452&0.090&0.209&175&12&5\\
TYC 9420-426-1&188.335188&-79.357768&11.35&12.51&G9IV/V&5362&0.135&0.153&194&15&5\\
TYC 9420-411-1&188.378563&-79.754723&9.04&9.57&F7V&...&-0.060&0.021&121&3&1\\
TYC 9420-72-1&188.589597&-79.677933&10.70&11.32&F8&...&0.000&0.064&220&12&2\\
TYC 9420-112-1&189.044929&-79.527371&11.48&13.01&K6V&4306&0.039&0.239&79&2&5\\
TYC 9420-68-1&189.067068&-79.526283&11.16&12.59&K3V&4934&0.225&0.162&77&2&5\\
TYC 9424-191-1&189.710111&-80.673848&10.76&11.48&G6V&5563&-0.095&0.068&143&6&6\\
TYC 9424-491-1&189.785193&-81.699802&11.07&11.80&G7V&5530&-0.098&0.081&236&43&5\\
TYC 9424-245-1&189.956631&-80.889986&9.39&10.90&K3IV&4505&0.263&0.058&245&18&5\\
TYC 9421-168-1&191.466754&-79.462328&8.64&9.21&F7V&...&-0.030&0.018&135&5&1\\
TYC 9425-804-1&191.499319&-81.679202&10.91&11.91&G8IV/V&5465&0.049&0.079&222&16&5\\
TYC 9425-717-1&191.657332&-80.902973&10.11&10.59&F8&...&-0.121&0.045&184&8&2\\
TYC 9425-608-1&191.747790&-81.264530&11.22&12.16&G3V&5702&0.136&0.120&195&13&5\\
TYC 9421-183-1&192.003101&-80.439730&11.57&12.69&K4IV/V&4546&-0.156&0.185&198&11&5\\
TYC 9425-1095-1&192.036601&-80.687633&11.54&13.36&G9IV/V&5344&0.692&0.293&194&23&6\\
TYC 9421-1019-1&192.172120&-79.780663&10.22&12.13&K3IV&4596&0.611&0.113&214&22&6\\
TYC 9425-398-1&192.195763&-81.688287&11.56&12.03&K4V&4599&-0.702&0.169&234&24&5\\
TYC 9425-1069-1&192.383889&-80.702354&7.22&7.43&A3/5IV&...&0.059&0.015&178&9&1\\
TYC 9421-1284-1&192.892038&-79.025840&10.30&11.05&F8&...&0.108&0.053&164&16&2\\
TYC 9425-443-1&193.185072&-81.870645&10.05&10.18&A0V&...&0.111&0.031&623&250&1\\
TYC 9425-976-1&193.310927&-80.972750&10.00&10.55&F9V&5995&-0.085&0.044&233&15&5\\
TYC 9421-713-1&193.656483&-80.072813&10.07&10.87&G2V&5752&0.034&0.048&93&2&5\\
TYC 9421-1213-1&193.866040&-80.000583&11.64&13.46&K2IV/V&5012&0.634&0.304&144&6&6\\
TYC 9421-704-1&194.036907&-79.432958&10.79&11.54&K2V&5040&-0.256&0.086&204&21&5\\
TYC 9425-19-1&194.463133&-80.651948&10.88&11.56&G8V&5451&-0.158&0.081&144&23&5\\
TYC 9421-683-1&194.928284&-79.946768&8.13&8.21&A1V&...&0.028&0.017&170&8&1\\
\end{longtable}
}

\begin{figure}
\resizebox{\hsize}{!}{\includegraphics{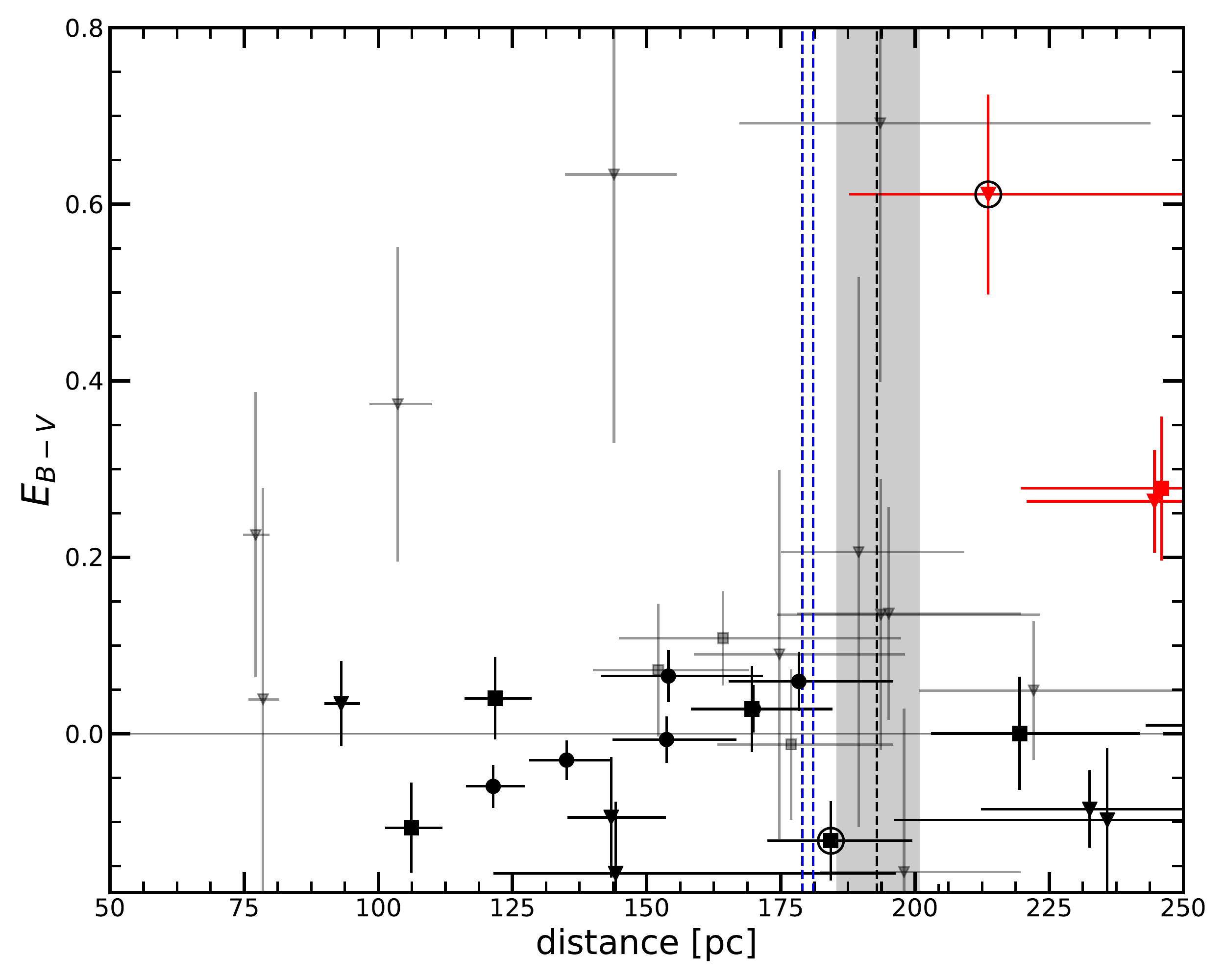}}
\caption{Color excess versus distance for the stars located on the line-of-sight of the Chamaeleon~III cloud. The black dashed line represents the position of the cloud front edge, while the blue dashed line is the distance to the Chamaelon~I and II clouds. Other symbols are as in Fig.~\ref{fig:E_BV_d_ChamI}.}
\label{fig:E_BV_d_ChamIII}
\end{figure}

We compute the color excess for all these stars and show it versus their distances in Fig. \ref{fig:E_BV_d_ChamIII}. This allows us to locate the distance of the front edge of the cloud at a parallax of 5.19$\pm$0.21 mas, corresponding to a distance of:
\begin{equation}
\rm d_{Chamaeleon\ III} = 193 ^{+8+12}_{-7-11} \ \textrm{[pc]}. \nonumber
\end{equation} 
Compared to the distances to the other clouds, this value is to be considered slightly more uncertain due to the higher level of structuring of this cloud (see Fig. \ref{fig:scat_chaIII}). It is possible that some denser cloudlets are closer, which makes the analysis more uncertain\footnote{We note that the well known T~Cha star is projected between Chamaeleon~III and Chamaleon~I in the sky, but it is actually in foreground at $d$=107 pc.}. Furthermore, a smaller number of stars are available for the analysis. In any case, this value represents the first ever estimate of the distance of this cloud, and confirms that its distance is compatible with the one of the other two clouds, although the nominal value is $\sim$10-15 pc further away. This result suggests that it is possible that these three Chamaeleon clouds are higher density star-forming regions of a single larger cloud structure.


\subsection{Musca}\label{sec:Musca}

\begin{figure}
\resizebox{\hsize}{!}{\includegraphics{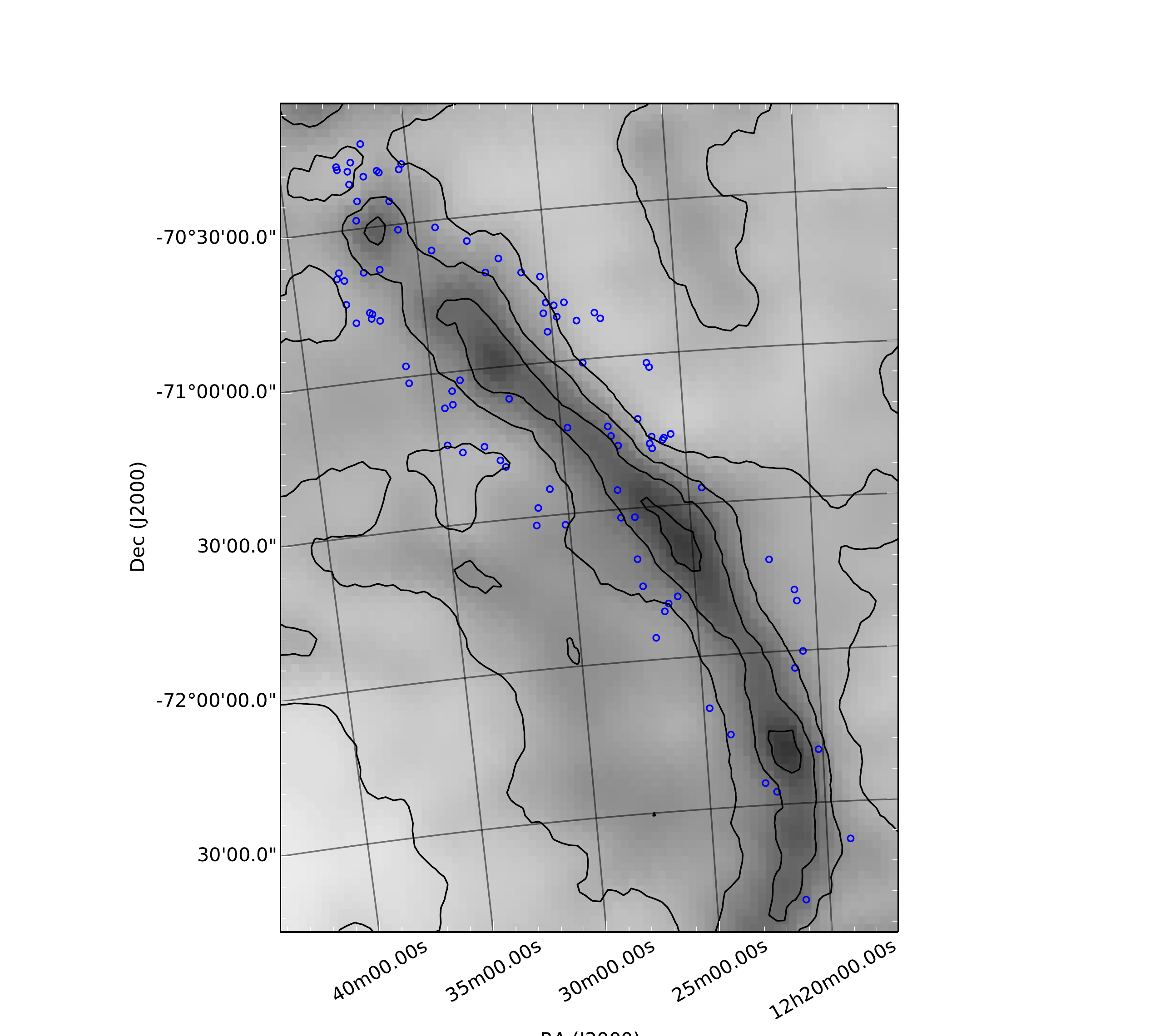}}
\caption{IRAS maps of Musca region in 100 $\mu$m flux. The six contours range from 8MJy/sr to 30 MJy/sr. Blue circles are the TGAS stars in this field of view.}
\label{fig:scat_Musca}
\end{figure}

The Musca cloud is an elongated and much narrower structure in the sky than the Chamaeleon clouds. We show the extent of this cloud in Fig. \ref{fig:scat_Musca}, together with the position of the 76 stars included in the TGAS catalog that are selected to determine the reddening turn-on distance for this cloud. For only six of these stars both spectral type and luminosity classes are reported in the literature, while for the remaining 70 stars only the effective temperature is available, and we derive their luminosity class as discussed in Appendix \ref{ap:CMDs}. We find that 20 of these stars are dwarfs, one is in between sub-giant and dwarf, 22 are sub-giants, and 27 are giants (see Table \ref{tab:Musca_stars}).

\longtab{
\begin{longtable}{llllllllllll}
\caption{\label{tab:Musca_stars} Catalog of stars on the line of sight of the Musca cloud used in this research}\\
\hline\hline
Name & Ra (ICRS) & Dec (ICRS) & V$_t$ & B$_t$ & SpT & T$_{\rm eff}$ [K] & E$_{(B-V)}$ & $\sigma_{E_{(B-V)}}$ & d [pc] & $\sigma_{d}$ & Ref.\\
\endfirsthead
\caption{continued.}\\
\hline\hline
Name & Ra (ICRS) & Dec (ICRS) & V$_t$ & B$_t$ & SpT & T$_{\rm eff}$ [K] & E$_{(B-V)}$ & $\sigma_{E_{(B-V)}}$ & d [pc] & $\sigma_{d}$ & Ref.\\
\endhead
\hline
\multicolumn{12}{l}{\tablebib{

- Spectral type reference : (1) \citet{1975mcts.book.....H}.

- Effective Temperature reference : (2) \citet{2006ApJ...638.1004A}; (3) \citet{2013AJ....146..134K}.

- $V_t$ and $B_t$ are taken from \citet{2000A&A...355L..27H}.

- The distance are taken from \citet{AABJ} based on TGAS.}}
\endfoot
Field Stars&&&&&&&&&&& \\ \hline 
TYC 9235-75-1&184.744341&-72.623027&10.98&11.64&F8IV&6104&0.039&0.117&343&49&2\\
TYC 9235-46-1&185.253211&-72.056947&11.22&11.76&K2V&5042&-0.436&0.140&150&9&2\\
TYC 9235-679-1&185.265659&-72.815646&9.33&9.85&F6V&...&-0.039&0.485&131&4&1\\
TYC 9235-735-1&185.519706&-72.458277&11.65&12.48&G9V&5365&-0.071&0.205&169&9&2\\
TYC 9235-880-1&185.978470&-72.262064&10.99&13.09&K1IV&4890&0.788&0.278&565&158&2\\
TYC 9236-2654-1&186.391663&-71.266527&11.59&11.90&G0III&5645&-0.391&0.200&1627&1231&2\\
TYC 9236-2708-1&186.440226&-71.798832&11.46&12.21&G7IV&5278&-0.212&0.155&777&243&2\\
TYC 9236-2694-1&186.477595&-71.283939&11.72&13.10&G0III&5645&0.510&0.281&1587&1385&2\\
TYC 9236-2833-1&186.540153&-71.819673&11.41&13.95&K3III&4250&0.877&0.384&1677&1205&3\\
TYC 9232-1139-1&186.560104&-71.043156&10.71&12.38&K3III&4250&0.139&0.194&1216&885&3\\
TYC 9236-2852-1&186.586967&-71.270860&11.69&12.13&F3IV&6657&0.002&0.204&757&640&2\\
TYC 9236-2551-1&186.590825&-71.309106&12.33&13.77&F5III&6339&0.766&0.385&1749&1554&2\\
TYC 9236-2559-1&186.612172&-71.292938&11.17&12.57&G9III&4788&0.224&0.157&1005&532&3\\
TYC 9236-2765-1&186.698257&-71.928683&11.04&12.60&G9III&4698&0.360&0.217&833&588&3\\
TYC 9232-987-1&186.714633&-71.209541&9.55&10.61&G8IV/V&...&0.161&0.738&169&7&1\\
TYC 9236-2625-1&186.793877&-71.757364&11.24&12.86&K0III&4661&0.378&0.220&907&537&3\\
TYC 9236-2475-1&186.822400&-71.529577&12.20&13.97&K7V&4088&0.172&0.396&161&8&2\\
TYC 9236-2525-1&186.829035&-71.667462&9.15&10.78&G8/K0III&...&0.406&0.981&504&70&1\\
TYC 9236-2787-1&186.934221&-71.291496&12.99&13.35&K5V&4491&-0.821&0.534&1171&731&2\\
TYC 9236-2618-1&186.965502&-71.527611&11.62&14.25&K1IV&4888&1.236&0.377&881&478&3\\
TYC 9236-2828-1&186.977664&-71.436525&11.37&12.20&M1.5V&3614&-0.782&0.246&241&15&2\\
TYC 9236-2608-1&186.998691&-71.258229&11.49&12.01&G8IV&5234&-0.444&0.147&843&311&2\\
TYC 9232-1571-1&187.009469&-70.871527&11.61&13.20&K4IV&4283&0.237&0.292&642&221&2\\
TYC 9232-1149-1&187.023990&-71.226137&10.75&12.91&K3IV&4453&0.816&0.190&588&381&3\\
TYC 9232-319-1&187.063387&-70.851113&11.88&14.04&K6IV&4000&0.589&0.364&866&605&2\\
TYC 9232-1377-1&187.222035&-71.011565&10.33&10.69&A9V&7416&0.051&0.187&596&385&2\\
TYC 9232-1021-1&187.432929&-71.218857&11.69&12.60&F2IV&6810&0.420&0.180&391&46&3\\
TYC 9232-331-1&187.461001&-70.815895&11.61&12.45&K4III&4156&-0.706&0.364&1937&1277&2\\
TYC 9232-659-1&187.538311&-70.804380&11.81&13.35&K1III&4576&0.212&0.313&2321&1668&3\\
TYC 9236-2491-1&187.542600&-71.534846&11.62&12.64&G0IV&6039&0.282&0.199&320&34&3\\
TYC 9232-1017-1&187.546179&-70.900162&12.19&12.75&K6V&4322&-0.783&0.345&402&64&2\\
TYC 9236-2877-1&187.669550&-71.413901&11.01&11.65&K5V&4459&-0.588&0.144&245&19&2\\
TYC 9232-1247-1&187.752816&-70.698897&11.20&12.10&G2IV&5887&0.123&0.124&609&133&2\\
TYC 9236-2502-1&187.806206&-71.471769&12.02&12.94&K2IV&4782&-0.146&0.251&757&303&2\\
TYC 9236-2580-1&187.839039&-71.528530&11.43&12.23&K1IV&4871&-0.317&0.151&643&379&2\\
TYC 9232-106-1&187.964050&-70.646002&9.46&10.81&G8/K0III&...&0.164&0.981&274&22&1\\
TYC 9232-1370-1&187.993974&-71.106936&12.54&13.70&K2III&4500&-0.186&0.347&1548&1372&3\\
TYC 9236-2603-1&188.094876&-71.327661&11.94&14.21&K3III&4250&0.652&0.423&1536&1207&3\\
TYC 9232-1237-1&188.105148&-70.687457&12.31&13.71&K5IV/V&4485&0.035&0.357&340&110&2\\
TYC 9236-2531-1&188.144294&-71.304669&11.78&12.56&G2IV&5845&0.011&0.214&657&167&2\\
TYC 9236-2832-1&188.291355&-71.254813&11.80&12.74&G9III&4781&-0.169&0.212&1153&785&2\\
TYC 9236-2539-1&188.516373&-71.266192&11.37&13.71&F4III&6512&1.565&0.360&1033&1405&2\\
TYC 9232-1331-1&188.552552&-70.523517&10.85&11.65&G2IV&5841&0.034&0.085&248&18&2\\
TYC 9232-1126-1&188.558231&-71.062732&12.40&13.01&F4III&6438&0.101&0.326&2513&1813&2\\
TYC 9232-729-1&188.565687&-71.106925&11.61&13.97&K4IV&4318&0.897&0.351&878&1349&3\\
TYC 9232-320-1&188.611316&-70.597344&11.51&13.32&K5III&4000&0.042&0.285&1915&1347&2\\
TYC 9232-884-1&188.649860&-71.115855&9.35&11.04&G8/K0III&...&0.460&0.981&391&53&1\\
TYC 9232-1173-1&188.663429&-71.237581&11.02&11.61&G4V&5673&-0.178&0.092&263&18&2\\
TYC 9232-247-1&188.812822&-70.305215&11.73&12.19&K1V&5136&-0.458&0.199&413&53&2\\
TYC 9232-1129-1&188.843248&-70.321599&9.31&9.68&F2V&6771&-0.066&0.081&238&15&2\\
TYC 9232-1431-1&188.916064&-70.517922&11.63&13.45&M0IV&3800&0.134&0.296&433&125&3\\
TYC 9232-1012-1&188.970454&-70.421836&11.20&12.04&K2V&5042&-0.177&0.107&101&2&2\\
TYC 9232-350-1&188.979805&-71.021352&11.06&11.90&K3V&4749&-0.274&0.116&215&12&2\\
TYC 9232-1228-1&188.991943&-70.965114&12.15&12.33&K5V&4348&-0.985&0.434&927&370&2\\
TYC 9232-1582-1&189.038109&-70.324600&12.86&12.75&M7V&2650&-2.153&0.548&1277&1424&2\\
TYC 9232-811-1&189.057881&-70.318146&12.21&14.08&K5III&4000&0.084&0.377&1721&1382&3\\
TYC 9232-782-1&189.139606&-70.640857&11.79&12.68&G0V&5942&0.160&0.204&223&15&3\\
TYC 9232-236-1&189.184860&-70.224850&11.66&13.70&F5III&6337&1.273&0.308&1104&1182&2\\
TYC 9232-235-1&189.192594&-70.332111&11.32&12.74&K4III&4215&-0.215&0.233&1460&1010&3\\
TYC 9232-561-1&189.194959&-70.807272&12.09&12.88&K6V&4209&-0.582&0.269&295&25&2\\
TYC 9232-1392-1&189.263442&-70.782965&10.50&12.01&K3IV&4600&0.261&0.104&402&48&3\\
TYC 9232-1483-1&189.277034&-70.797565&11.23&11.30&F2III&6780&-0.297&0.235&1828&1493&2\\
TYC 9232-1584-1&189.281169&-70.410098&11.21&13.11&K7IV&3966&0.332&0.229&633&942&3\\
TYC 9232-1535-1&189.287324&-70.777418&11.09&11.51&F0III&7805&0.053&0.079&944&734&2\\
TYC 9232-1095-1&189.300499&-70.644492&11.01&11.52&K2V&4987&-0.454&0.095&141&5&2\\
TYC 9232-134-1&189.301922&-70.281244&11.40&12.88&K3III&4258&-0.019&0.247&1834&1280&3\\
TYC 9232-1166-1&189.312148&-70.472633&10.35&10.98&G6V&5576&-0.161&0.051&129&6&2\\
TYC 9232-400-1&189.339081&-70.352227&11.36&13.36&K4IV&4430&0.590&0.265&795&375&3\\
TYC 9232-73-1&189.340474&-70.309828&11.72&14.41&K5III&4000&0.788&0.419&1425&1192&3\\
TYC 9232-1287-1&189.431963&-70.805618&11.84&13.38&K2III&4509&0.147&0.300&2202&1658&3\\
TYC 9232-430-1&189.438833&-70.300674&13.11&13.19&M6V&2842&-1.935&0.648&622&153&2\\
TYC 9232-1430-1&189.444197&-70.290485&12.48&14.38&M4III&3546&-0.022&0.813&2218&1622&2\\
TYC 9232-1341-1&189.499866&-70.663582&8.59&8.78&B9III/IV&...&0.218&0.063&615&113&1\\
TYC 9232-633-1&189.508909&-70.741924&10.79&12.28&K3IV&4463&0.246&0.141&675&203&3\\
TYC 9232-1521-1&189.543697&-70.636117&11.61&13.15&K1III&4532&0.208&0.273&1065&819&3\\
TYC 9232-1592-1&189.569260&-70.655453&12.12&12.41&F0III&7715&-0.062&0.207&2055&1514&2\\ 
\end{longtable}
}

\begin{figure}
\resizebox{\hsize}{!}{\includegraphics{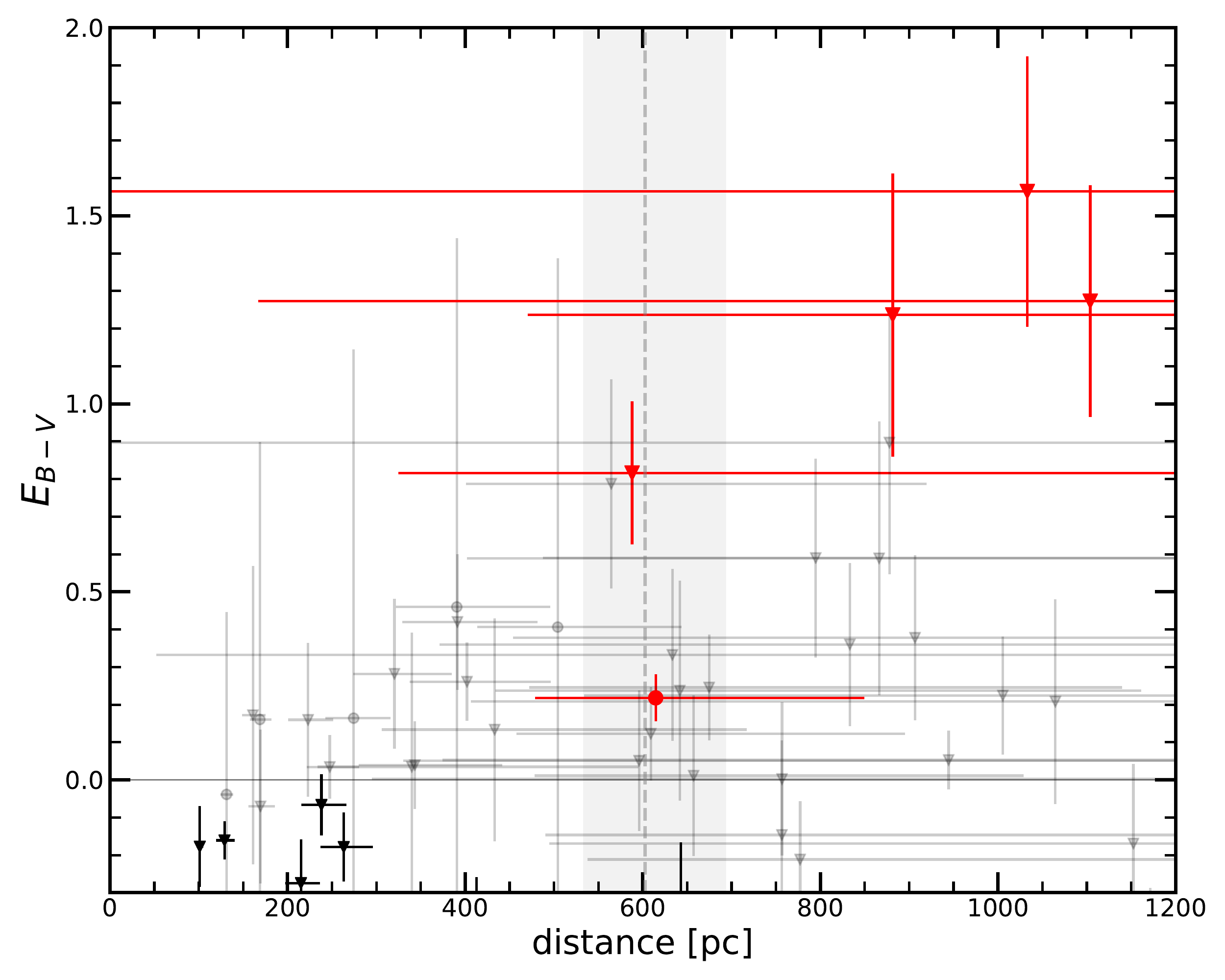}}
\caption{Extinction distribution along the line of sight of Musca. Symbols as in Fig.~\ref{fig:E_BV_d_ChamI}. The dashed line is the inverted mean value of the parallaxes of the first two reddened stars, with its statistical uncertainty shown with a gray region.}
\label{fig:E_B_V_Musca}
\end{figure}

The analysis of the color excess versus distance for these 76 stars, shown in Fig. \ref{fig:E_B_V_Musca}, does not allow us to constrain the distance of the front edge of the Musca cloud. We can only infer the mean parallax of the two closest significantly reddened stars, which is found to be 1.66$\pm$0.22, corresponding to a distance of $603^{+91+133}_{-70-92}$ pc. This can be considered as an upper limit on the distance to this cloud. 
The impossibility to derive a more stringent constraint on the distance to this cloud is due to the large uncertainties on the color excess of the stars on the line of sight, mainly due to the fact that only the effective temperature is available for the vast majority of these stars, and that these temperatures are in many case photometrically computed, thus more uncertain. 


\subsection{Summary}\label{sec:res_sum}

We report the final distances to the four clouds analyzed here in Table \ref{tab:Sum_res}. The distances to the Chamaeleon I is found to be $\sim$20 pc larger than previously assumed, although the difference is still compatible within the measurement uncertainties. On the contrary, the Chamaeleon~II cloud is confirmed to be at a distance of $\sim$180 pc, and is found to be almost at the same distance as the Chamaeleon~I cloud. 

We are able to report for the first time a distance estimate to the Chamaeleon~III cloud, which is consistent with the idea that the three Chamaeleon clouds are part of a single cloud structure, and significantly larger than the previously assumed distance to this cloud. Finally, the distance to the Musca cloud is only loosely constrained to be closer than 600 pc, with a large uncertainty.


\begin{table}
\caption{Summary of the distances to the clouds studied here}\label{tab:Sum_res}
\centering
\begin{tabular}{lc}
\hline\hline
Name & Distance [pc]\\\hline
Chamaeleon I & $179 ^{+11+11}_{-10-10}$ \\
Chamaeleon II & $181 ^{+6+11}_{-5-10}$ \\
Chamaeleon III & $193 ^{+8+12}_{-7-11}$ 
\end{tabular}
\end{table}

\section{Discussion}\label{sec:Discussion}

\subsection{Impact of the new distance estimate on stellar properties of Chamaeleon~I stars}\label{sec:Diff_discussion}

As noted in the introduction, the distance is a key parameter in the computation of several stellar properties, in particular the stellar luminosity ($L_{\star}$). This value is used together with the effective temperature and evolutionary models to derive the stellar mass ($M_{\star}$) and age. We have determined that the distance to the Chamaeleon~I region is $179 ^{+11+11}_{-10-10}$ pc, and not 160$\pm$15 pc as previously assumed. Although the two values are compatible within their uncertainties, here we explore the impact of the difference in the nominal value of the distance on the stellar properties of the young stellar objects in this region. 

\begin{figure}
\resizebox{\hsize}{!}{\includegraphics{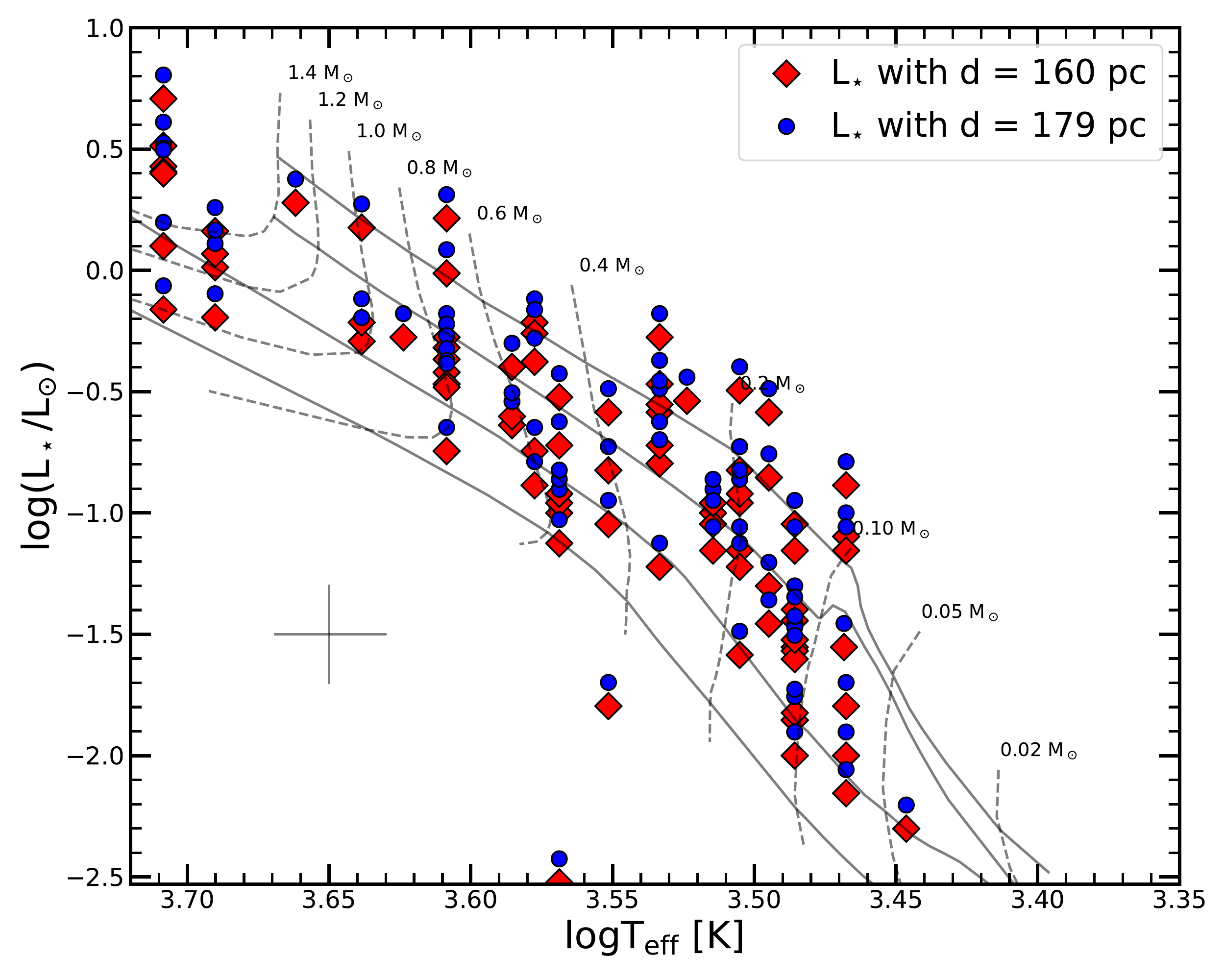}}
\caption{Hertzsprung Russel diagram (HRD) of the young stars with disks in the Chamaeleon I region studied by \citet{2016A&A...585A.136M,2017arXiv170402842M}. The red diamonds and blue circles are used for $L_{\star}$ computed with the distance of 160 pc found by \citet{Whittet_1997} and using our distance of 179 pc, respectively. Typical uncertainties are shown with a cross in the bottom left. The evolutionary models (isochrones in solid lines and evolutionary tracks with dashed lines) are taken from \citet{2015A&A...577A..42B}.}
\label{fig:HRD_ChaIm}
\end{figure}

We use in the following the values of $L_\star$ determined with spectroscopy and assuming a distance of 160 pc by \citet{2016A&A...585A.136M,2017arXiv170402842M}. The stellar luminosity is modified by a factor equal to the squared ratio of the new and previous distance estimate ${d^2_{179pc}}/{d^2_{160pc}}$. This implies that $L_\star$ is higher by 25\% due to the larger distance estimate. 
The difference in the values of $L_\star$ is shown on the Hertzsprung-Russel diagram (HRD) in Fig. \ref{fig:HRD_ChaIm}. 
The values of $L_\star$ obtained using the newly derived distance to the Chamaeleon~I region are still compatible within 1$\sigma$ with those using the previously used distance. 

In turn, the mass and the age of the stars are also affected by the different value of the distance to the region, and we compute the new values using the evolutionary models by \citet{2015A&A...577A..42B}. The variations on the stellar masses are negligible, since this value mainly depends on the temperature of the stars for low mass stars at this evolutionary stage (cf. Fig.~\ref{fig:HRD_ChaIm}) , which is unaffected by the distance estimate. On the contrary, the age estimate using the new distance suggests that the Chamaeleon~I stellar population is $\lesssim 1$ Myr younger than previously found.


\subsection{The connection between the Chameleon clouds and large scale structures}\label{sec:geo_sheet}

Now that the distance of the three Chamaeleon clouds is constrained we put this information together with the position of these clouds in galactic coordinates to investigate the connection between these clouds and large-scale structures in the Milky Way galaxy. 

\begin{figure}
\resizebox{\hsize}{!}{\includegraphics{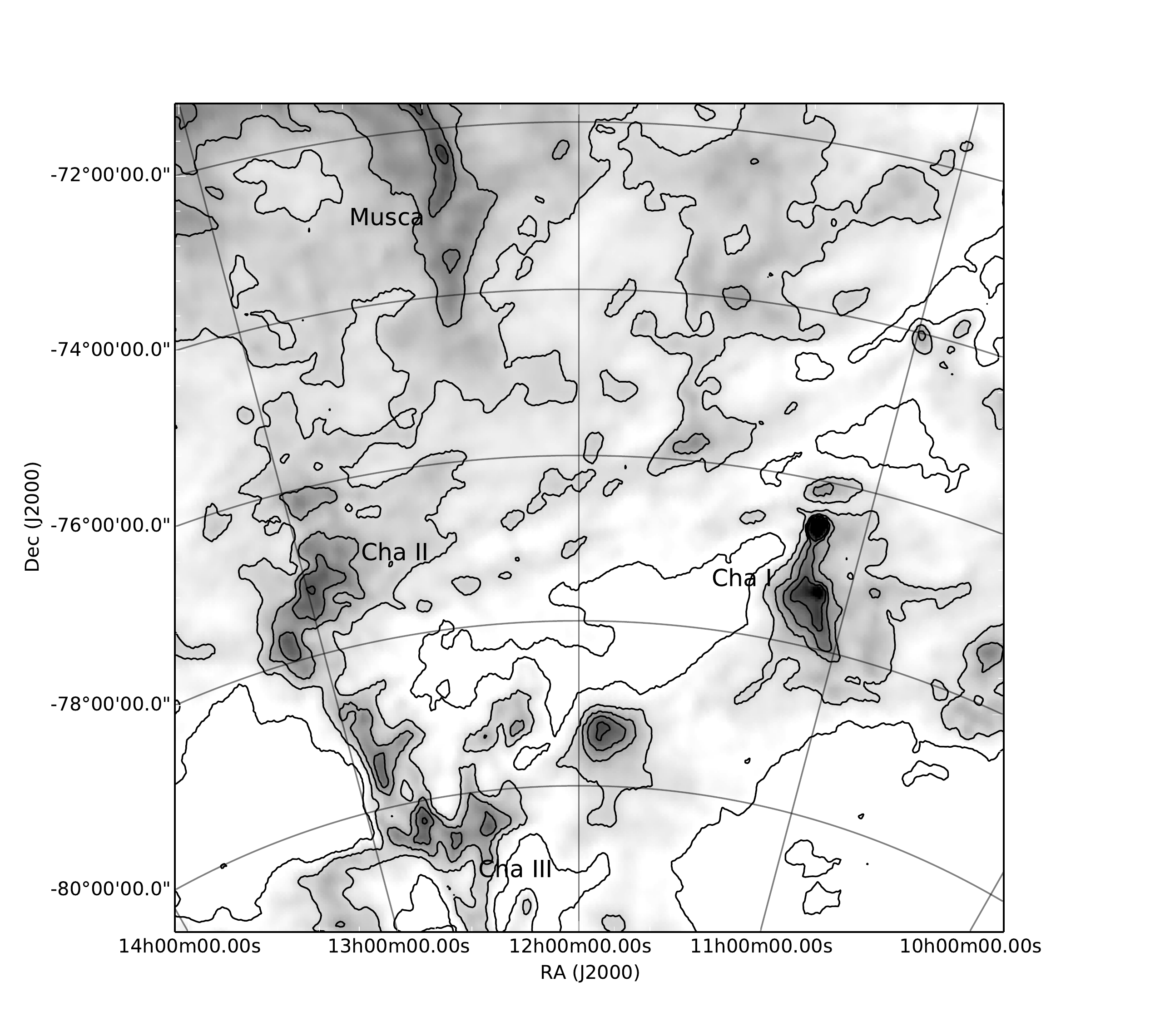}}
\caption{IRAS 100$\mu$m flux map of Chamaeleon-Musca region. The six contours range from 8MJy/sr to 30 MJy/sr.}
\label{fig:Cha_reg_img}
\end{figure}

We have shown that the Chamaeleon~I and Chamaeleon~II clouds are both at $\sim$180 pc from us, while the Chamaeleon~III region is slightly further away. According to \cite{1997A&A...326.1215C}, there is negligible interstellar material between us and the Chamaeleon-Musca complex, as observed by the low extinction measured. Thus, emission from dust particles should arise from clouds located at the same distance or further away than the Chamaeleon-Musca complex. We show in Fig. \ref{fig:Cha_reg_img} a dust density map of this complex, which shows the presence of dust structures connecting the three Chamaeleon clouds with each other and with the Musca cloud. This structure could represent a physical link between the clouds.

We discuss the connection between this structure and larger scale ones in the context of two hypotheses reported in the literature. 
 \citet{King01101979} observed the nebulosity structures in the vicinity of the southern Celestial Pole and 
proposed the existence of a sheet-like structure of clouds extending in parallel to the galactic plane and southwards from it, with a distance of between 40 and 80 pc from the plane.
On the contrary, \citet{1997A&A...326.1215C} investigated the color excess as a function of distance in a large region of the sky ranging from the Southern Coalsack (l=$\rm 300^{\circ}$,b=$0^{\circ}$) to the Chamaeleon-Musca complex (l=$\rm 300^{\circ}$,b=$-15^{\circ}$), They found evidence of a possible physical association between these two structures, and thus hypothesized the existence of a sheet-like structure of clouds perpendicular to the galactic plane 
They suggested that these clouds could be over-densities formed at the interface between the Local and Loop I bubbles. 

\begin{figure}
\resizebox{\hsize}{!}{\includegraphics{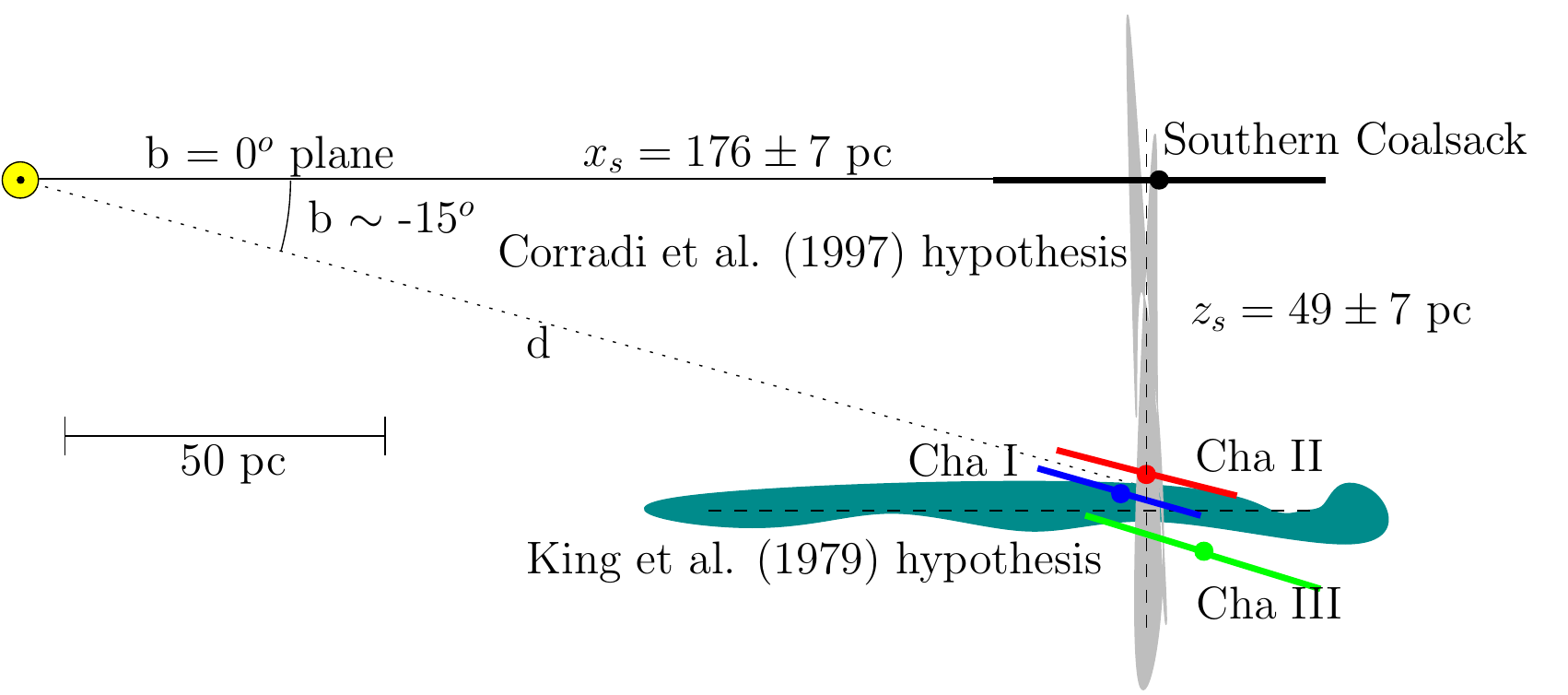}}
\caption{Schematic view of the large scale structure of the dust clouds in the direction of the Chamaeleon-Musca complex. The location of the Sun is represented with a yellow circle, while the blue, red, green, and black points represent the location of the three Chamaeleon clouds and southern Coalsack dark cloud, respectively, derived using the distances from our work in the case of the Chamaeleon clouds, and the galactic coordinates of the objects. The large gray and green cloud are the dust and gas hypothetical structure proposed by \citet{King01101979} and \citet{1997A&A...326.1215C}, respectively, as discussed in the text.}
\label{fig:Schema_geometry}
\end{figure}

We represent our results and these two hypotheses with a scheme in Fig. \ref{fig:Schema_geometry}. The distance to the sheet of clouds parallel to the galactic plane ($z_s$) and the one to the sheet of clouds perpendicular to it ($x_s$) are obtained using the following equations:
\begin{equation}
z_s = d \cdot sin(|b|) \ \mathrm{and} \ x_s = d \cdot cos(|b|),
\label{eq::dist}
\end{equation}
where $d$ is the distance and $b$ the galactic latitude of the clouds.
 
\begin{table}
\caption{$\rm z_s \ and \ x_s$ for the studied clouds}\label{tab:Sum_res}
\centering
\begin{tabular}{lccc}
\hline\hline
Name & Gal. Latitude &$z_s$ [pc] &  $x_s$ [pc]\\\hline
Chamaeleon I & $-16 \pm 1$ & $\rm 49 \pm 4$ & $\rm 172 \pm 13$\\
Chamaeleon II & $-14 \pm 1$ & $\rm 44 \pm 3$ & $\rm 176 \pm 14$\\
Chamaeleon III & $-17 \pm 1$ & $\rm 56 \pm 6$ & $\rm 185 \pm 21$\\\hline
 &Mean & $\rm 49 \pm 7$ & $\rm 176 \pm 7$
\end{tabular}
\end{table}

We can determine the distance from the galactic plane to the sheet of clouds parallel to it \citep{King01101979}, and the one from the Sun to the sheet of clouds perpendicular to the galactic plane \citep{1997A&A...326.1215C} using the distance to the three Chamaeleon dark clouds. The values obtained with the distance to each of the clouds are reported in Table \ref{tab:Sum_res}. All the three estimates of the distance of the hypothetical parallel sheet of clouds from the galactic plane are consistent with each other within 2$\sigma$, at most, and similarly are the distances from the Sun of the perpendicular sheet of clouds. Thus, we cannot distinguish between the two hypotheses solely based on the measured distances to the three Chamaeleon clouds. 


An independent information on the distance to the southern Coalsack region leads us to consider the hypothesis of a sheet of clouds perpendicular to the galactic plane a viable possibility. This region is located at galactic latitude $b=0^{\circ}$, and thus $x_s$ corresponds to its distance (see Eq.~\ref{eq::dist}). The distance to this region was measured by \cite{1960MNRAS.120..163R} and \cite{1989A&A...215..119F} to be $174 \pm 18$ pc and $180 \pm 26$ pc, respectively. These values are well compatible with the values of $x_s$ derived here using the distance to the Chamaleon~I and II clouds, which supports the hypothesis of the existence of a sheet of clouds perpendicular to the galactic plane. However, we cannot with our data discard either of the two hypothesis, or that both structures exist.


Finally, we try to put more stringent constraints on the distance to the Musca cloud in the context of the two hypotheses of large scale structures we discussed. By knowing the galactic latitude of this region ($b$ = -9 deg) and assuming the values of $x_s$ and $z_s$ of Table~\ref{tab:Sum_res} we can invert Eq.~\ref{eq::dist} to infer a distance to the Musca cloud. The result of this exercise is a distance $d_{\rm parallel} = 311 \pm 54$ pc or $ d_{\rm perpendicular} = 179 \pm 7$ pc according to the two hypotheses, respectively. Both values are well consistent with the loose upper limit we have derived from the color excess analysis.

\section{Conclusion}\label{sec:Conclusion}

In this work we have used the newly released data from the Data Release 1 of the Gaia satellite, in particular the parallax estimates from the TGAS catalog, to constrain the distance to the four main clouds in the Chamaeleon-Musca complex. Due to the limiting magnitude of the TGAS catalog we were able to derive a distance by directly computing the mean value of the distances to members only in the Chamaeleon~I cloud. This analysis of the distances to the members led us to estimate a distance of $188 ^{+6+11}_{-6-10}$ pc for this cloud, where the errors are the statistical and systematic one, respectively.

We then combine the parallax values from the TGAS catalog for stars on the line of sight of the Chamaeleon-Musca clouds with measurements of their $E(B-V)$ color excess. By measuring the parallax at which we observe an increase of the color excess to values larger than zero we are able to locate the front edge of the cloud, and thus to infer its distance. This method allows us to measure a distance of $179 ^{+11+11}_{-10-10}$ pc to the Chamaeleon~I cloud, which is compatible with the distance to the members and further away than the previously assumed distance to this cloud. The reason for the difference is a difference in the photometric distance estimates with respect to the TGAS distances, which shows the importance of parallax based distances for the analysis.
Furthermore, we confirm the distance to the Chamaeleon~II cloud with the reddening turn-on method, and measure a distance of $181 ^{+6+11}_{-5-10}$ pc, consistent with the one to the Chamaeleon~I cloud. We are then able to measure for the first time the distance to the Chamaeleon~III cloud, which is found to be $193 ^{+8+12}_{-7-11}$ pc, and only to suggest that the distance to the Musca cloud is smaller than $603^{+91+133}_{-70-92}$ pc. 


The newly estimated further distance to the Chamaeleon~I cloud results in an increase of the stellar luminosities of 25\%, which implies a slightly younger age for this region by $\lesssim$1 Myr, but it affects the other stellar properties only within their uncertainties.



Finally, we have discussed the relation of the clouds in the Chamaeleon-Musca region with two different possible large scale cloud structures hypothesized in the literature, namely a sheet of cloud parallel or perpendicular to the galactic plane. The distances we measured to the three Chamaeleon clouds, combined with their galactic coordinates, do not allow us to rule out either of the hypotheses. In the context of these hypotheses, we compute two possible distances to the Musca cloud, namely $\rm d_{parallel} = 311 \pm 54$ pc and $\rm d_{perpendicular} = 179 \pm 7$ pc. 


The major limitation of our study is represented by the limited number of members of the targeted clouds included in the TGAS catalog, mainly due to the limiting magnitude of this catalog. This limitation will be largely overcome by the future Gaia Data Release 2, planned for April 2018. Parallaxes for stars as faint as $G \sim 20$ magnitudes will be included, and will allow us to use the distance to a larger number of members to derive the distances to nearby star forming regions ($d\lesssim$300 pc). On the other hand, we will be able to apply the reddening turn-on method to star forming clouds even in the cases where the members of the region are too faint to be included in the Gaia catalog, for example where most of their members are still embedded in their envelope or where star formation has not started yet.

\begin{acknowledgements}
This work has made use of data from the European Space Agency (ESA)
mission {\it Gaia} (\url{http://www.cosmos.esa.int/gaia}), processed by
the {\it Gaia} Data Processing and Analysis Consortium (DPAC,
\url{http://www.cosmos.esa.int/web/gaia/dpac/consortium}). Funding
for the DPAC has been provided by national institutions, in particular
the institutions participating in the {\it Gaia} Multilateral Agreement. This research has made use of the SIMBAD database,
operated at CDS, Strasbourg, France. JV acknowledges the ESA trainee program and CFM acknowledges an ESA research fellowship funding. We acknowledge a detailed and constructive anonymous referee report which significantly improved the paper. Finally, we would like to thank Jos De Bruijne, Eleonora Zari, and Stella Reino for their stimulating comments.
\end{acknowledgements}

\bibliographystyle{aa} 
\bibliography{aa} 

\begin{appendix}

\section{Color Magnitude Diagrams to determine luminosity classes of stars}
\label{ap:CMDs}

Our method of deriving distances to dark clouds from the determination of the reddening turn-on distance relies on the availability of spectral type, luminosity class, photometry, and distance for each star on the line-of-sight. Here we describe how we derive the luminosity class for the targets when this is not available in the literature. We convert the observed $V$ magnitude of the targets into absolute magnitude using the parallax estimates from the TGAS catalog converted into distances as discussed by \citet{AABJ}. Here we use the standard deviation on the mode of the posteriors as an uncertainty on the distance. This value is well in line with using the 5 and 95 percentile of the posteriors for small parallax errors, as in the majority of the stars used here. Small differences in these uncertainties causing a misclassification do not affect the results significantly, as we discuss in the following. We use this information together with the $B-V$ color to locate the stars on the color-magnitude diagram (CMD). Then, we compare the location of the stars on the CMD with the color and magnitudes reported for different luminosity classes by \citet{2013ApJS..208....9P} for dwarfs (V) and K1-M5 sub-giants (IV), and by \citet{Gottlieb_1978} for earlier type sub-giants, for giants (III), and for luminosity classes I and II. 

\begin{figure}
\resizebox{\hsize}{!}{\includegraphics{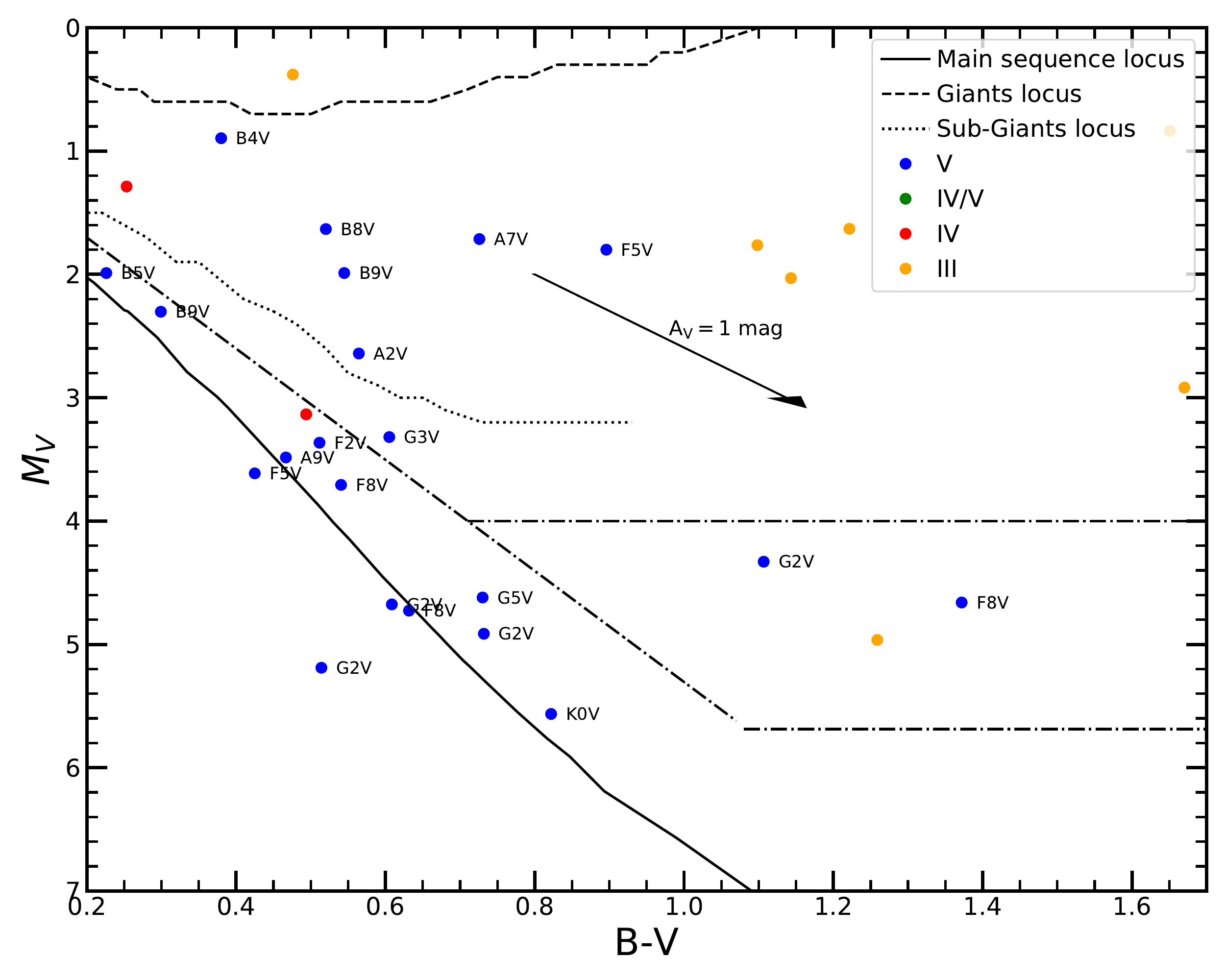}}
\caption{Absolute $V$ magnitude versus $B-V$ color for the stars on the line-of-sight of the Chamaeleon~I cloud with known spectral type and luminosity class, as labeled. The black line represents the Main Sequence locus, the dashed line represent the Giants locus and the dotted line the Sub-Giants locus \citep[][see text for details]{2013ApJS..208....9P,Gottlieb_1978}. The dashed and dotted line represent the limit between luminosity class V, IV/V and IV. The black arrow is the reddening vector.}
\label{fig:CMD_control}
\end{figure}

Fig. \ref{fig:CMD_control} shows the CMD for stars on the line-of-sight of the Chamaeleon~I cloud which have both spectral type and luminosity class reported in the literature. This diagram confirms that the luminosity class estimated in the literature agrees with the loci expected for the different luminosity classes, with only a few outliers. We note, however, that a misclassification in the luminosity class between sub-giants and dwarfs would have a limited effect on the color excess value. Indeed, the difference in $E(B-V)$ between a B8V star and a B8IV star is of 0.018 mag, and similar for other spectral types. This uncertainty is included in the error budget and does not affect substantially our analysis. Finally, we note that the reddening vector is parallel to the dwarf and sub-giant locus, as shown on Fig. \ref{fig:CMD_control}, thus increasing values of extinction do not affect our results.



\begin{figure}
\resizebox{\hsize}{!}{\includegraphics{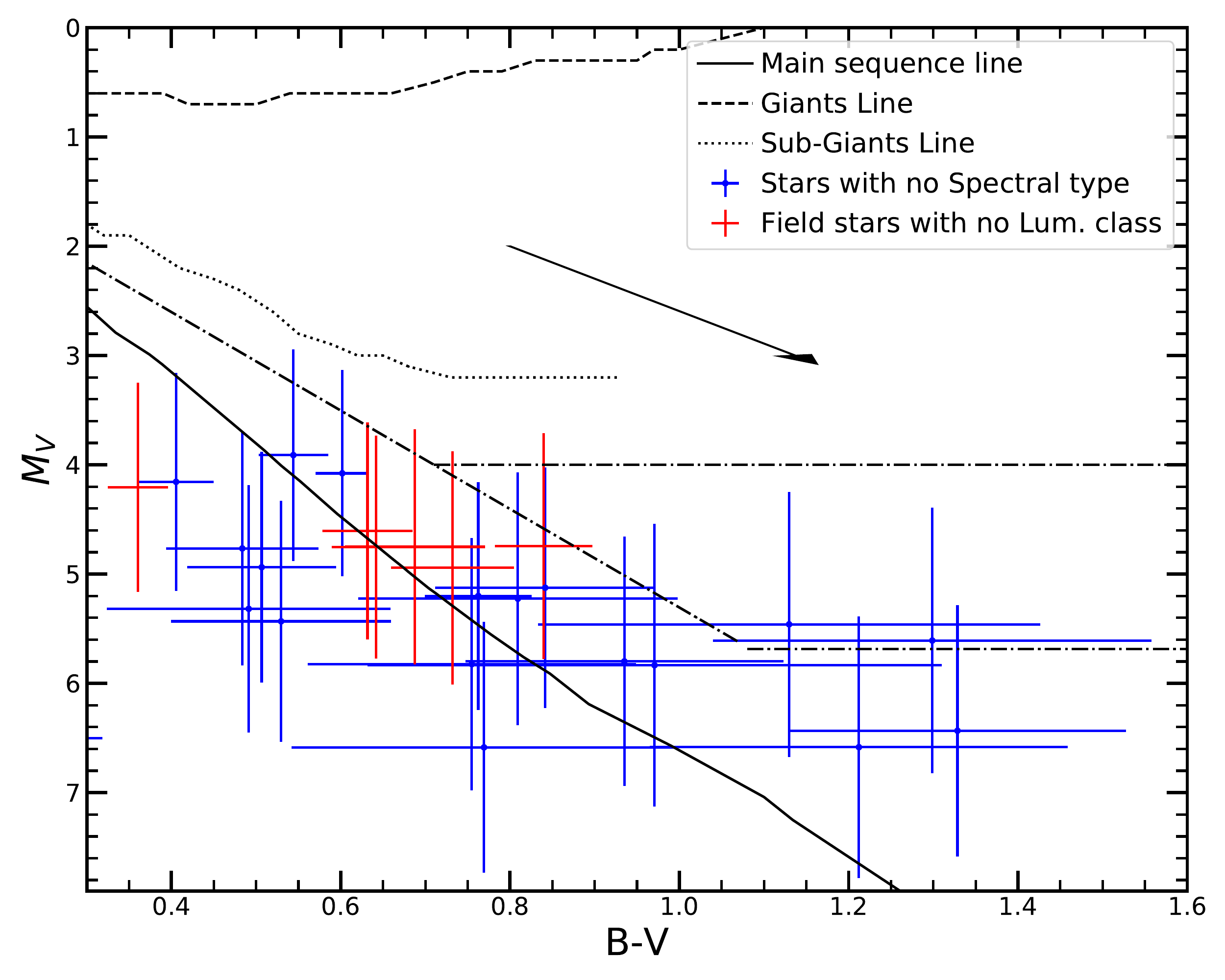}}
\caption{Absolute $V$ magnitude versus $B-V$ color for the stars on the line-of-sight of the Chamaeleon~I cloud without an estimate of the luminosity class. Red symbols are used for stars with spectral type derived from spectroscopy and missing the luminosity class, while blue symbols are used when only the effective temperature is available. Lines are as in Fig.~\ref{fig:CMD_control}.}
\label{fig:CMD_ChaI}
\end{figure}

\begin{figure}
\resizebox{\hsize}{!}{\includegraphics{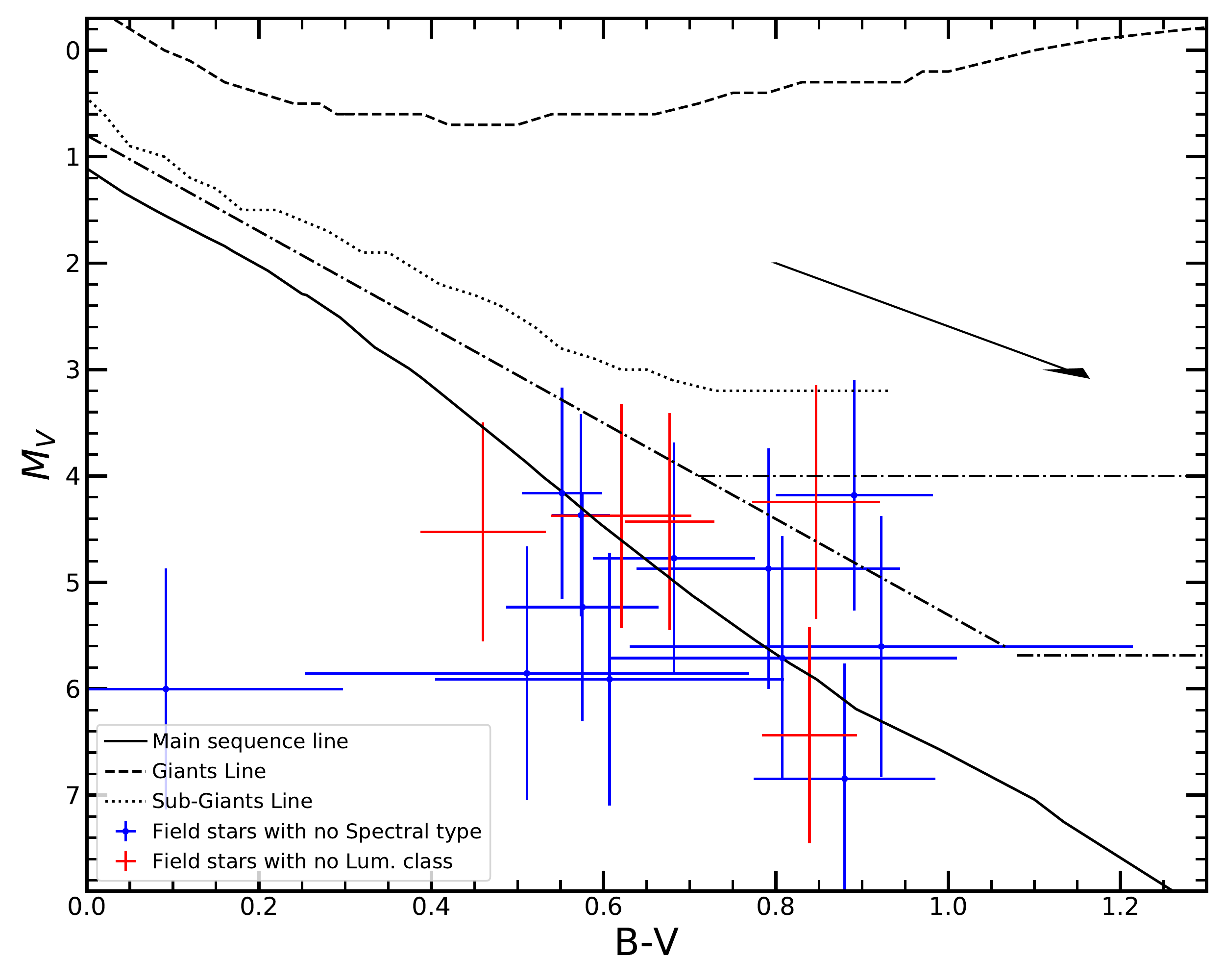}}
\caption{Same as Fig.~\ref{fig:CMD_ChaI} for stars on the line-of-sight of the Chamaeleon~II cloud.}
\label{fig:CMD_ChaII}
\end{figure}

\begin{figure}
\resizebox{\hsize}{!}{\includegraphics{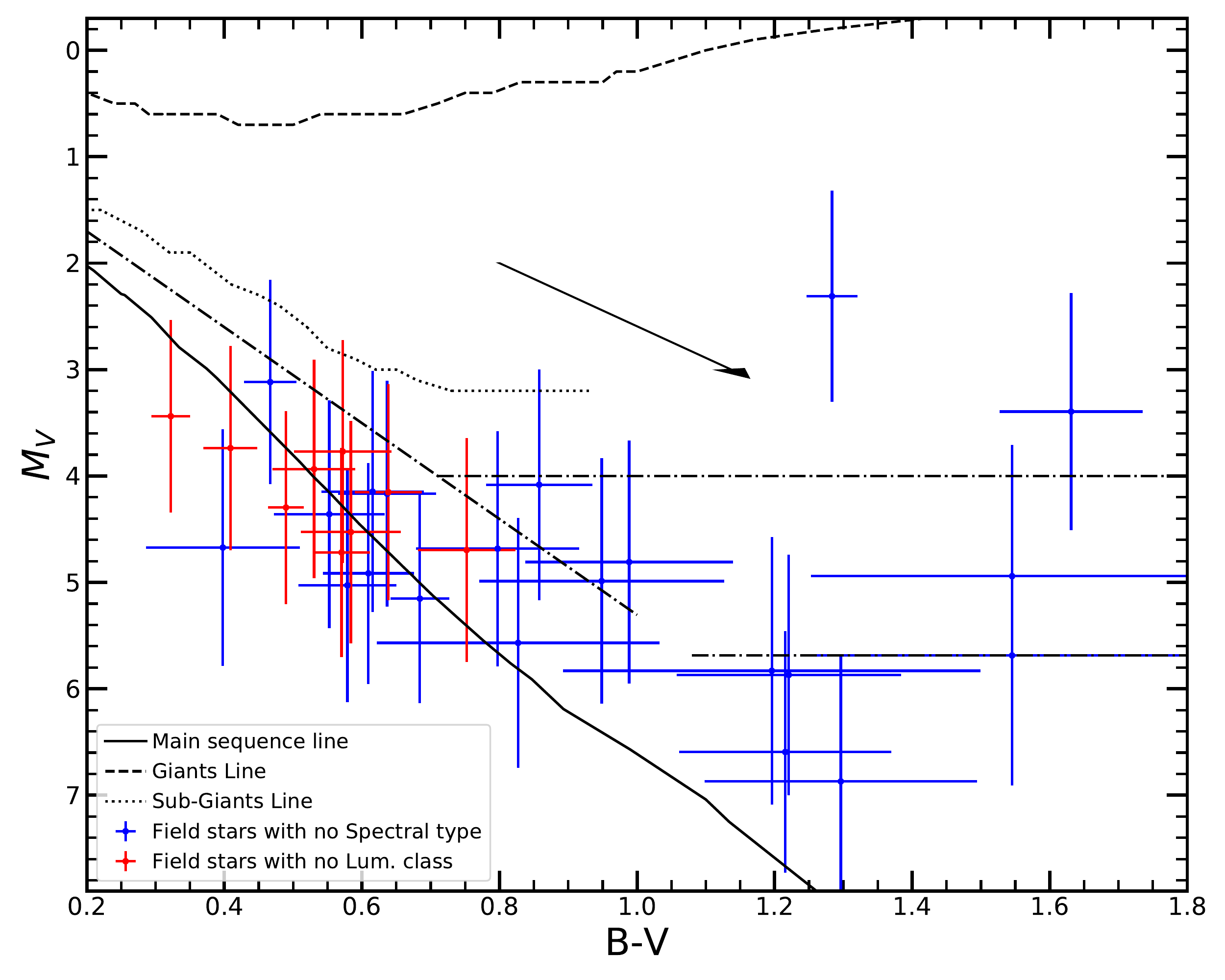}}
\caption{Same as Fig.~\ref{fig:CMD_ChaI} for stars on the line-of-sight of the Chamaeleon~III cloud.}
\label{fig:CMD_ChaIII}
\end{figure}

\begin{figure}
\resizebox{\hsize}{!}{\includegraphics{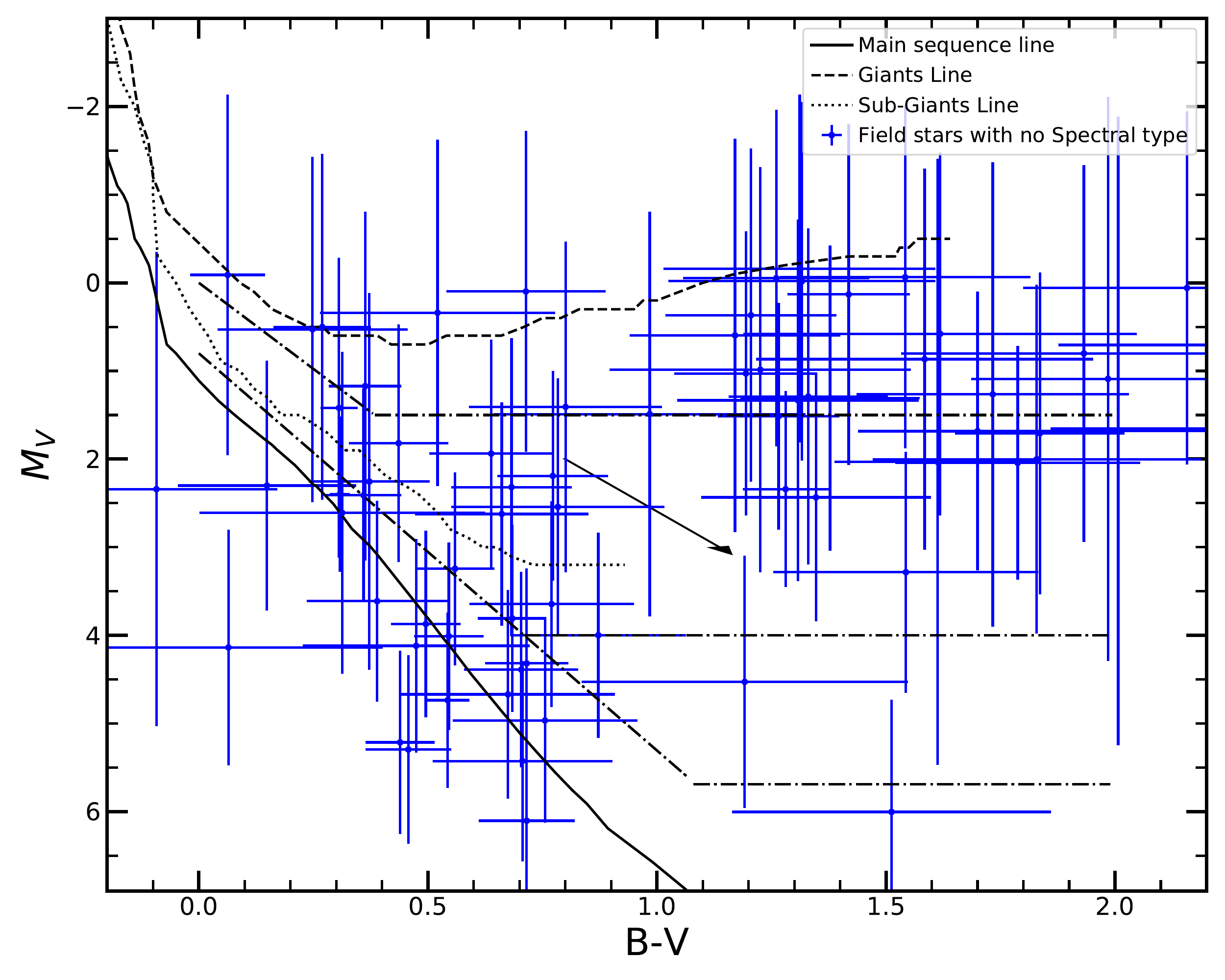}}
\caption{Same as Fig.~\ref{fig:CMD_ChaI} for stars on the line-of-sight of the Musca cloud.}
\label{fig:CMD_Musca}
\end{figure}

The CMDs used to determine the luminosity class for stars on the line of sight of the four clouds analyzed here are shown in Figs. \ref{fig:CMD_ChaI}, \ref{fig:CMD_ChaII}, \ref{fig:CMD_ChaIII} and \ref{fig:CMD_Musca} for the Chamaeleon I, II, III, and Musca clouds, respectively. Most of the stars in the line-of-sight of the Chamaeleon~I and II have luminosity class V, with only two exceptions in each region. In the Chamaleon~III line of sight, instead, seven stars are not of luminosity class V, and an higher fraction of luminosity class III and IV is present on the line-of-sight of the Musca cloud. 


\section{Discrepancies between photometric distances and TGAS distances}
\label{ap:discr_Dist}

\begin{figure}
\resizebox{\hsize}{!}{\includegraphics{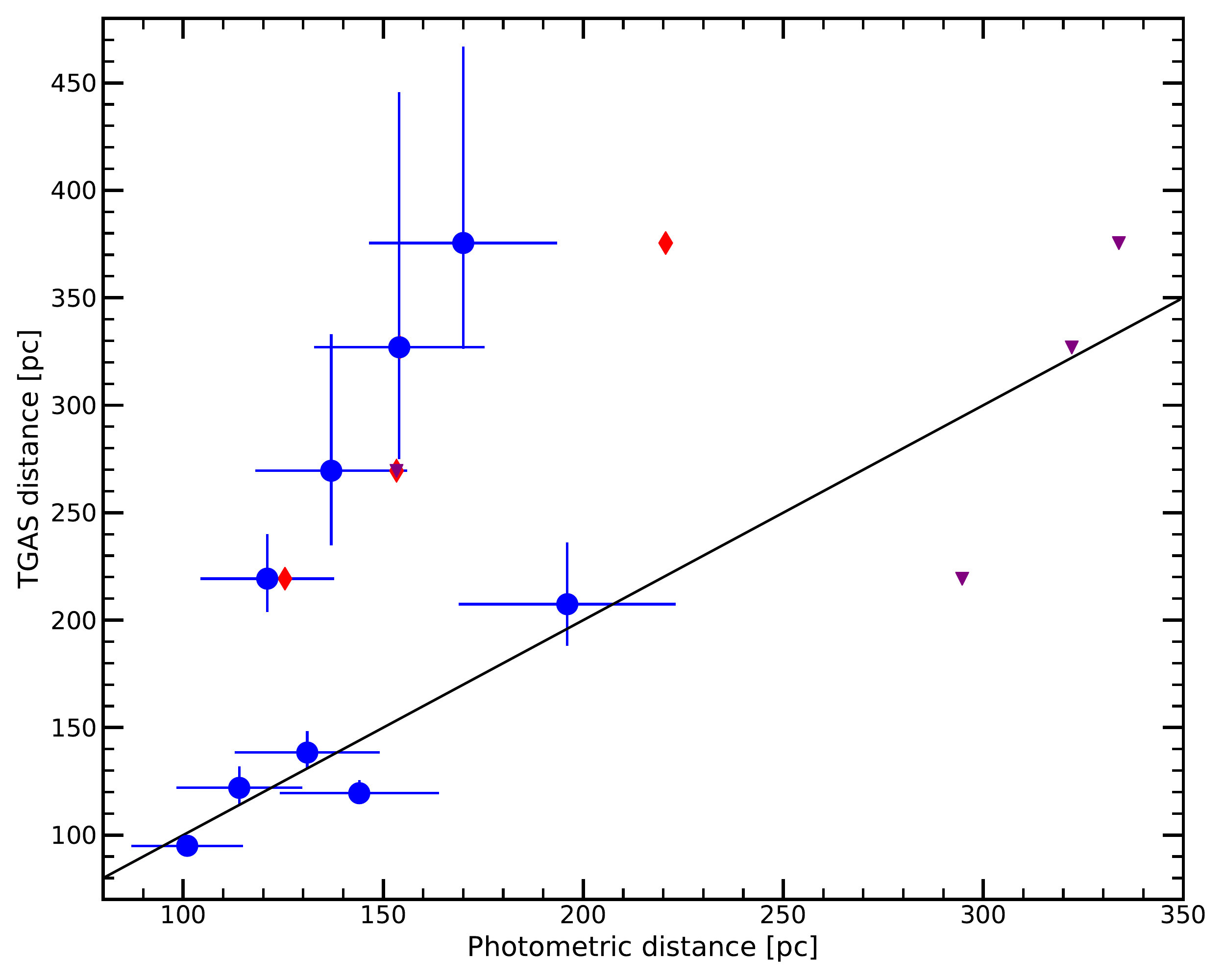}}
\caption{Distances estimated from TGAS parallax measurements versus the photometric distances used by \citet{Whittet_1997} for the same stars, shown with blue points. Red diamonds are used to show the photometric distance computed assuming $ R_V = 3.1$, and purple triangles the same value correcting the luminosity class of the targets using their position on the CMD in Fig. \ref{fig:CMD_TGAS_phot_Dist}.}
\label{fig:TGAS_phot_Dist_Comp_corrected}
\end{figure}

The previous estimate of the distance to the Chamaeleon~I cloud of 160$\pm$15 pc was obtained by  \citet{Whittet_1997} using the same reddening turn-on method as used by us. Here we investigate the origin of the different estimate, which is related to the difference in the distances estimated from the TGAS parallax versus the photometric distances used by \citet{Whittet_1997}, shown in Fig. \ref{fig:TGAS_phot_Dist_Comp_corrected} for the stars in common in the two works. These values are reported in Table~\ref{tab:comp_dist}. 
While there is an agreement between the two methods for four stars with distances smaller than 150 pc and for one star at $\sim$200 pc, there are four additional stars  (TYC 9414-642-1, TYC 9410-2627-1, TYC 9410-2608-1, 9411-998-1) for which the two methods do not agree, with the TGAS distances being significantly larger. These stars were considered to be in front of the cloud by \citet{Whittet_1997} based on the photometric distances, while they appear to be behind the cloud with the TGAS estimates. 

We first explore the possibility that the differences are due to the reddening law assumed by \citet{Whittet_1997} and used to compute the photometric distance. They assumed a value of $R_V = 1.1 \times E(V-K)/E(B-V)$, which results in very different values from one target to another. We thus assume $R_V=3.1$ for all the stars and recalculate their photometric distances. As shown in Fig.~\ref{fig:TGAS_phot_Dist_Comp_corrected}, the differences in the photometric distances are too small to explain the discrepancy between the two methods.

\begin{figure}
\resizebox{\hsize}{!}{\includegraphics{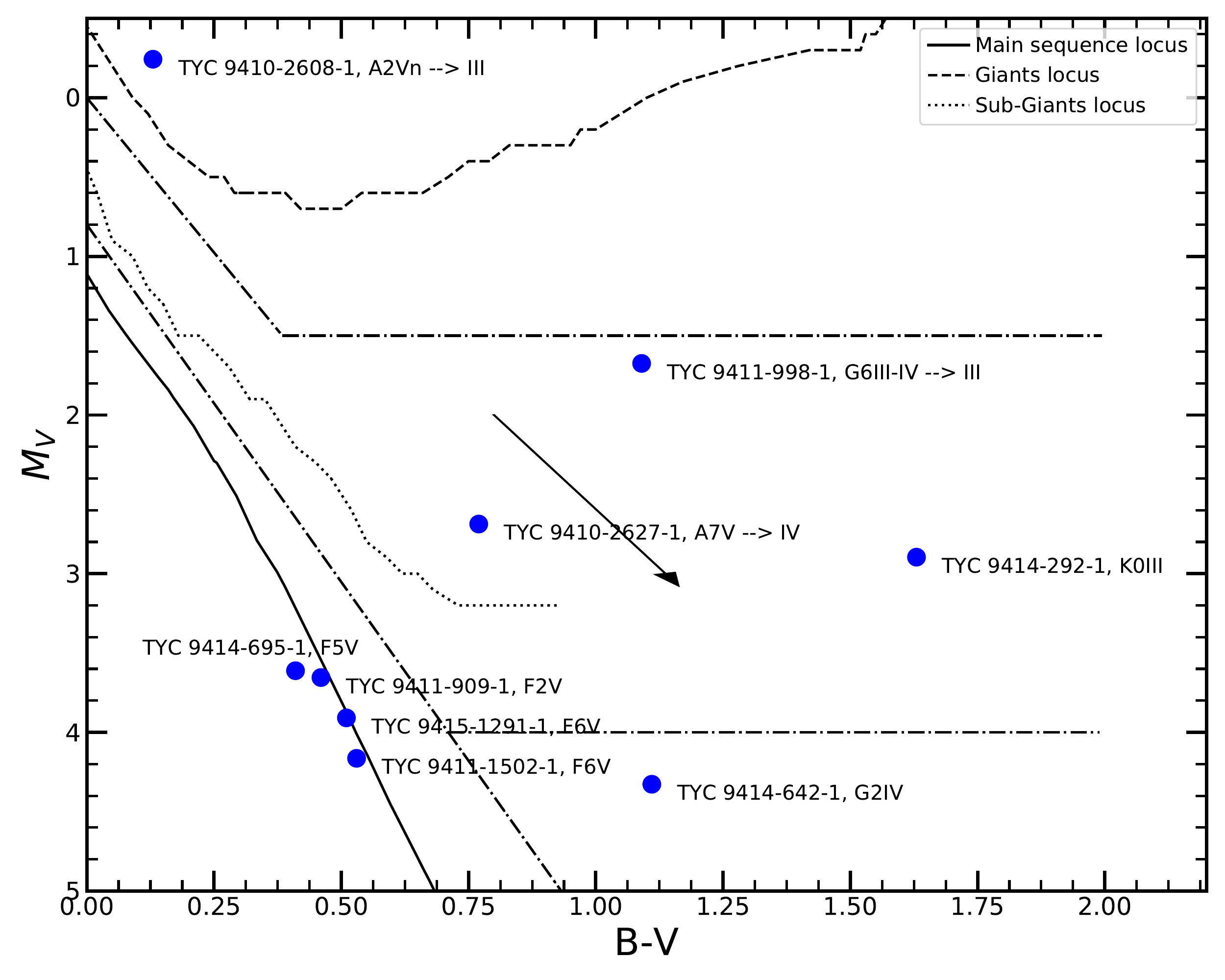}}
\caption{CMD of the stars in Table \ref{tab:comp_dist}. The absolute magnitude are computed using the TGAS distances and the $B$ and $V$ are taken from \citet{2000A&A...355L..27H}. The Tycho 2 id and spectral classification \citep{Whittet_1997} of each star is reported and the lum. class corresponding to the position on the CMD computed with the TGAS distance.}
\label{fig:CMD_TGAS_phot_Dist}
\end{figure}

We thus consider the hypothesis that the luminosity class of the stars was incorrectly estimated, and for this aim we show in Fig.~\ref{fig:CMD_TGAS_phot_Dist} the CMD of the stars in Table~\ref{tab:comp_dist}, where the absolute $V$ magnitudes are computed using the TGAS distances. 
The four F-type dwarfs and the K-type giant are the five stars for which the distances derived with the two methods are compatible. Also their positions on the CMD is always consistent with the luminosity class assumed by \citet{Whittet_1997}. Similarly, the G-type sub-giant is still consistent with the previously assumed luminosity class, and thus the discrepancy between the photometric distance and the TGAS distance cannot be explained with this argument. On the opposite, we find that the luminosity class was incorrectly estimated for the remaining three stars with discrepancy in the distances estimates with the two methods. 
Once we compute the photometric distances with the correct luminosity class and assuming $R_V=3.1$ (Fig.~\ref{fig:TGAS_phot_Dist_Comp_corrected}) we find a better agreement between this distance and the TGAS distances.

We conclude that the reason for the different estimate of the distance to the Chamaeleon~I cloud is due to the incorrect estimate of the photometric distance of four stars in the work by \citet{Whittet_1997}, which lead the authors to derive a closer distance to this region. The reason for the erroneously estimated photometric distances is due to a wrong luminosity class determination for these targets. Once correcting for this error, the agreement between the TGAS and photometric distances is good. This shows the importance of having distances estimates derived from parallaxes, as these are independent on the stellar parameters.


\begin{table}
\caption{Objects used in the comparison}
\label{tab:comp_dist}
\centering
\begin{tabular}{l c c c}
\hline\hline
Tycho 2 Id & $\mathrm{d_{phot}}$ [pc] & $\mathrm{d_{TGAS}}$ [pc] \\
\hline
TYC 9414-642-1 & 137 & 270 \\
TYC 9410-2627-1 & 170 & 375 \\
TYC 9410-2608-1 & 154 & 327 \\
TYC 9411-998-1 & 121 & 219 \\
TYC 9414-292-1 & 196 & 207 \\
TYC 9415-1291-1 & 114 & 122 \\
TYC 9411-909-1 & 131 & 138 \\
TYC 9414-695-1 & 101 & 95 \\
TYC 9411-1502-1 & 144 & 119\\
\hline
\end{tabular}
\tablebib{Photometric distance are taken from \citet{Whittet_1997} and the TGAS distances are taken from \cite{AABJ}}
\end{table}

\end{appendix}

\end{document}